\documentclass[fleqn,usenatbib]{mnras}

\usepackage{mathptmx}
\usepackage[T1]{fontenc}

\DeclareRobustCommand{\VAN}[3]{#2}
\let\VANthebibliography\thebibliography
\def\thebibliography{\DeclareRobustCommand{\VAN}[3]{##3}\VANthebibliography}

\usepackage{graphicx}	% Including figure files
\usepackage{amsmath}	% Advanced maths commands
\usepackage{amssymb}	% Extra maths symbols
\usepackage{gensymb}

\usepackage[dvipsnames]{xcolor}
\usepackage{ulem}

\title[Searching for nascent planetary nebulae]{Searching for nascent planetary nebulae: OHPNe candidates in the SPLASH survey}

\author[R.~A. Cala et al.]{
Rold{\'a}n A. Cala$^{1}$\thanks{E-mail: rcala@iaa.es (RC)}, 
Jos{\'e} F. G{\'o}mez$^{1}$,
Luis F. Miranda$^{1}$, 
Lucero Uscanga$^{2}$, 
Shari L. Breen $^{3}$,
Joanne R. Dawson$^{4}$,
\newauthor
Itziar de Gregorio-Monsalvo$^{5}$,
Hiroshi Imai$^{6,}$$^{7}$,
Hai-Hua Qiao $^{8,}$$^{9}$,
Olga Su{\'a}rez$^{10}$
\\
$^{1}$Instituto de Astrof{\'i}sica de Andaluc{\'i}a, CSIC, Glorieta de la Astronom{\'i}a s/n, E-18008 Granada, Spain\\
$^{2}$Departamento de Astronom{\'i}a, Universidad de Guanajuato, A.P. 144, Guanajuato 36000, Gto., Mexico\\
$^{3}$SKA Observatory, Jodrell Bank, Lower Withington, Macclesfield, SK11 9FT, UK \\
$^{4}$Department of Physics and Astronomy and MQ Research Centre in Astronomy, Astrophysics and Astrophotonics, Macquarie University, NSW 2109, Australia\\
$^{5}$European Southern Observatory (ESO), Alonso de C{\'o}rdova 3107, Vitacura, Santiago 763-0355, Chile\\
$^{6}$ Amanogawa Galaxy Astronomy Research Center, Graduate School of Science and Engineering, Kagoshima University, 1-21-35 Korimoto, Kagoshima,\\ Kagoshima 890-0065, Japan\\
$^{7}$ Center for General Education, Institute for Comprehensive Education, Kagoshima University, 1-21-30 Korimoto, Kagoshima,  Kagoshima 890-0065, Japan\\
$^{8}$National Time Service Center, Chinese Academy of Sciences, Xi'An, Shaanxi, China, 710600\\
$^{9}$Shanghai Astronomical Observatory, Chinese Academy of Sciences, 80 Nandan Road, Shanghai, China, 200030\\
$^{10}$ Laboratoire Lagrange, Observatoire de la C{\^{o}}te d’Azur, Universit{\'e} C{\^{o}}te d’Azur, CNRS, Bd de l’Observatoire, CEDEX 4, 06304 Nice, France \\
}

\date{Accepted XXX. Received YYY; in original form ZZZ}

\pubyear{2022}

\begin{document}
\label{firstpage}
\pagerange{\pageref{firstpage}--\pageref{lastpage}}
\maketitle

\begin{abstract}
The evolution of asymptotic giant branch stars from the spherical symmetry into the diverse shapes of planetary nebulae (PNe) is a topic of intensive research. Young PNe provide a unique opportunity to characterize the onset of this transitional phase. In particular, OH maser-emitting PNe (OHPNe) are considered nascent PNe. In fact, only 6 OHPNe have been confirmed to date. In order to identify and characterize more OHPNe, we processed the unpublished continuum data of the interferometric follow-up of the Southern Parkes Large-Area Survey in Hydroxyl (SPLASH). We then matched the interferometric positions of OH maser and radio continuum emission, considering the latter as a possible tracer of free-free emission from photoionized gas, characteristic of PNe. We report 8 objects with a positive coincidence, 4 of which are classified as candidate OHPNe here for the first time (IRAS 16372--4808, IRAS 17494--2645, IRAS 18019--2216 and OH 341.6811+00.2634). Available evidence strongly indicates that they are evolved stars, while the comparison with confirmed OHPNe indicates that they are likely to be PNe. Their final confirmation as bona fide PNe, however, requires optical/infrared spectroscopy. The obtained spectral indices of the radio continuum emission (between $\simeq$ 0.4 -- 1.3) are consistent with partially optically thick free-free emission from photoionized gas. Also, they cluster in the same region of a \textit{WISE} colour-colour diagram as that of the confirmed OHPNe ($9.5\la [3.4]-[22]\la 13.5$, and $4.0\la [4.6]-[12] \la 7.0$), thus this diagram could help to identify more OHPNe candidates in the future.

%The abstract should briefly describe the aims, methods, and main results of the paper.
%It should be a single paragraph not more than 250 words (200 words for Letters).
%No references should appear in the abstract.
\end{abstract}


%%%%%%%%%%%%%%%%%%%%%%%%%%%%%%%%%%%%%%%%%%%%%%%%%%

%%%%%%%%%%%%%%%%% BODY OF PAPER %%%%%%%%%%%%%%%%%%

\section{INTRODUCTION}

Stars with initial masses ($M$$_{\rm i}$) between 0.8 and 8 M$_{\odot}$ will undergo the phase to form a planetary nebula (PN) before becoming a white dwarf. PNe are conspicuous objects in optical images showing round, elliptical, bipolar, multipolar or point-symmetric morphologies \citep[e.g.,][]{IAC96, sah11} that are a result of the particular mass--loss history during the previous evolutionary stages: the asymptotic giant branch (AGB) and the post--AGB phases.

The AGB phase is characterized by strong mass--loss episodes, with rates up to 10$^{-4}$ M$_{\odot}$ yr$^{-1}$ \citep{vass93, block95}. This creates an expanding circumstellar envelope (CSE), whose kinematics is traceable with spectral lines at radio wavelengths. Maser line emission from different molecules has been widely used in these studies. In particular, stars with $M$$_{\rm i}$ = 0.8 -- 1.5 M$_{\odot}$ and $M$$_{\rm i}$ = 4 -- 8 M$_{\odot}$ are expected to develop an oxygen-rich (C/O < 1) CSE \citep{groene95}. In oxygen-rich AGBs, the CSE seems to be stratified with SiO, H$_{2}$O and OH masers located up to $\sim$10 AU, $\sim10^{2}$ AU and $\sim$10$^{4}$ AU from the central star, respectively \citep{reid81, eli92, dia94}. Thus, OH masers typically trace the front and back sides of the CSE \citep{reid76} with a typical terminal expansion velocity of $\sim$15 -- 30 km s$^{-1}$ \citep[][]{telin91}, whereas the H$_{2}$O and SiO ones trace inner regions that expand more slowly.

During a brief post--AGB transition period ($\sim 10^{2}-10^{4}$ yr) between the AGB and PN phases \citep[depending on $M$$_{\rm i}$;][]{block95, mil16}, the mass--loss rate drops down to $10^{-7}-10^{-8}$ M$_{\odot}$ yr$^{-1}$ \citep[][]{vass94}. The star shrinks at a constant luminosity, and its effective temperature ($T_{\rm eff}$) gradually increases, while the CSE continues expanding. With this sudden decline of mass--loss, the favorable conditions for maser-pumping are expected to disappear soon after the end of the AGB phase, on timescales of $\sim$10 yr, $\sim$100 yr and $\sim$1000 yr for SiO, H$_{2}$O and OH masers, respectively \citep{lew89, gom90}. 
When the surface temperature of the central star reaches $T_{\rm eff}$ $\simeq$ 25,000 -- 30,000 K, it emits enough ultraviolet (UV) photons to develop and sustain a photoionized nebula in its CSE. When this photoionized region is maintained, we can consider that the star is in the PN phase.
Molecules such as SiO, H$_{2}$O and OH are rapidly destroyed as the ionization front advances. Thus, comparing these maser lifetimes with the duration of the post--AGB phase, masers have been expected to be detected only in nascent PNe. However, special physical conditions in the CSE can temporarily shield these molecules from dissociation \citep{mir01} and maser-pumping can also be produced in collimated mass ejections with lower mass--loss rates than in the strong AGB winds \citep{sua09}. In particular, OH masers can trace bipolar outflows in the AGB and post--AGB phases \cite[]{chap88,zij01}, and so far, there are fifteen known cases where H$_{2}$O masers trace high-velocity collimated jets \citep[$\simeq 50-300$ km s$^{-1}$;][]{imai07,gom17}. These objects are called water fountains (WFs). 

Considering that maser emission is expected to disappear soon after the onset of the PN phase, it naturally follows that the number of maser-emitting-PNe must be small. So far, only 8 PNe have been confirmed to harbour maser emission, consistent with these short timescales of maser survival. Among these, 6 PNe display OH maser emission \citep[OHPNe;][]{usc12, qia16a} and 5 ones show H$_{2}$O maser emission \citep[H$_{2}$O-PNe;][]{mir01, deG04, gom08, usc14, gom15}, 3 of which harboring both species. SiO maser emission has never been detected in a PN \citep[e.g.][]{nym97, deg01}. With such scarcity, it is difficult to draw strong conclusions on the nature of maser-emitting PNe. Hence, detecting more of these sources is so crucial that the physical processes occurring at the beginning of photoionization could determine the subsequent morphological, kinematical and chemical evolution of evolved stars through the PN phase. Thus, further studies of these maser-emitting PNe can provide a deeper understanding about the end of low- and intermediate-mass stellar evolution. In principle, all confirmed maser detections in PNe trace O-rich CSE, although at least in one case \citep[the H$_{2}$O-PNe IRAS 18061--2505;][]{gom08} dual chemistry is present, with a C-rich central star and an O-rich circumstellar ring \citep[][and references therein]{mir21}. Another possible case of maser emission in dual-chemistry in a PN is the OH detection towards IRAS 07027--7934  \citep[e.g.,][]{zij91}, but the association of the maser with the PN requires confirmation with interferometric radio observations. The relationship of these peculiar PNe with other dual-chemistry evolved objects with maser emission in previous evolutionary phases \citep{ohn13} remains to be determined. Regarding C-rich stars, maser emission of molecules such as HCN is present at the AGB phase \citep{men18}, but so far, no such maser has been reported in PNe. Thus, as a consequence of their different chemical evolution, it has not been possible to use the presence of masers as a putative signpost of the youth of C-rich PNe.              

Focusing on the particular case of OH, \citet{dsp79} reported Vy 2--2 as the first possible case of PN with 1612 MHz maser emission, based on Arecibo observations. Afterwards, \citet{zij89} cataloged 12 OHPNe candidates, mostly based on single-dish observations. To date, only 6 PNe have been confirmed interferometrically to be associated with OH maser emission. These objects are Vy 2--2, IRAS 17393--2727, K 3--35, IRAS 17347--3139, JaSt 23, and IRAS 16333--4807 \citep{usc12, qia16a}. There is an additional PN, NGC 6302, that is known to be associated with OH emission, but it has been argued that its emission could be of thermal nature (not maser), since the CO and the 1720 MHz line profiles (the latter in absorption) are similar to that of OH at 1612 MHz \citep{gom16}, and that the 1612 MHz OH line is not detected at the longest baselines of ATCA \citep{qia20}. In this case, we cannot confidently include NGC 6302 in the group of bona fide OHPNe. 
OH masers in PNe seem to preferentially trace magnetized toroidal structures, perpendicular to the lobes of the nebulae. This is suggested in IRAS 17347--3139 \citep[based on the spatial distribution of the H$_{2}$O masers;][]{deG04, taf09}, and is observed in K 3--35 and IRAS 16333--4807 \citep[]{mir01, usc08, gom09, qia16a}. In Vy 2--2, IRAS 17393--2727 and JaSt 23, the spatial distribution of OH masers is still to be determined.

In order to increase the number of known maser-emitting PNe, we searched for OHPNe candidates in data from the interferometric follow-up observations of the Southern Parkes Large-Area Survey in Hydroxyl (SPLASH).

The SPLASH project \citep{daw22} performed a fully-sampled, high-sensitivity, unbiased survey of the four OH ground-state transitions ($^{2}\Pi_{3/2}$, $J=3/2$ at 1612.231, 1665.402, 1667.359 and 1720.530 MHz), covering 176 deg$^2$ of the southern Galactic plane and Galactic Center: $332^\circ \lid l \lid 10^\circ$, $|b|=2^\circ$ and $358^\circ \lid l \lid 4^\circ$, $ +2^\circ \lid b \lid +6^\circ$. The four OH transitions were observed with the 64-m radio telescope at the Parkes Observatory ($\sim$15 arcmin resolution). The scientific goals of SPLASH included both thermal and maser emission of OH. Regarding OH masers, in contrast to other surveys, which were targeted toward a sample of selected sources \citep*{chen93,deac04} and/or covered only one of the transitions \citep[e.g., the ATCA/VLA OH 1612 MHz survey of][]{sev97a, sev97b, sev01}, SPLASH allows a direct comparison among all four OH lines, while its large spatial coverage and high sensitivity provides an unbiased view of the nature and pumping mechanism of circumstellar masers.

Candidate maser sources detected at first in SPLASH were subsequently observed with the Australia Telescope Compact Array %\footnote{The Australia Telescope Compact Array is part of the Australia Telescope National Facility which is funded by the Australian Government for operation as a National Facility managed by CSIRO.} 
(ATCA), which allowed us to confirm unambiguously the OH maser detections, as well as an accurate determination of their positions. In this paper, we will refer to \cite{qia16b, qia18, qia20} interferometric follow-up maser observations as SPLASH--ATCA. 

In SPLASH--ATCA, 934 individual sources were found to produce OH maser emission in at least one of the ground-state transitions, 432 of which were sources detected for the first time. The detected objects were then classified depending on available literature, their infrared colours and compactness, the number of OH maser transitions and/or their spectral profiles: 629 evolved stars, 158 star forming regions (SFRs), 14 supernova remnants and 132 objects of unknown nature.

OHPNe candidates can be identified using radio continuum emission as a possible tracer of the presence of photoionized gas. Free-moving electrons in the ionized region of PNe produce bremsstrahlung (free-free) radiation that can be detected at radio wavelengths. Thus, cross-matching OH-masers with radio continuum sources can be used to identify candidate OHPNe \citep{zij89, usc12}. However, other objects, such as H\,{\sc ii} regions associated with massive young stellar objects (YSOs) can also show both radio continuum and OH maser emission. 

The ATCA correlator used in SPLASH--ATCA allows the simultaneous observation of line and continuum emission. \citet{qia16b, qia18, qia20} only reported the results of the OH line data. Therefore, we used the unpublished continuum data of SPLASH--ATCA as the main focus of this paper, in order to find positional coincidence between the OH-emitters and radio continuum sources, considering continuum emission as a tracer of photoionized gas in PNe.

We note that not all OH maser detections in SPLASH were followed-up with SPLASH--ATCA, since those sources already observed in the Mapping the Galactic Magnetic field through OH masers (MAGMO) project with ATCA \citep{green12} were not included. However, the MAGMO targets were most likely to be star-forming regions, so we do not think our data processing left out any evolved star covered by the original SPLASH survey.

\section{OBSERVATIONS AND DATA PROCESSING}

\subsection{SPLASH--ATCA continuum data}

\label{sec:obs_splash}

SPLASH--ATCA observations (project C2872) were carried out in several sessions from 2013-Oct-24 to 2016-Mar-04. The Compact Array Broadband Backend (CABB) correlator of ATCA was set in its 1M-0.5k mode, total bandwidth of 2 GHz centered at 2.1 GHz, and sampled over 2048 wideband channels of 1 MHz width each. These are used in this paper to obtain the continuum emission of the sources. Zoom windows were placed at several of these wideband channels to obtain line data at a higher spectral resolution, presented in \citet{qia16b, qia18, qia20}. We refer the reader to these papers for further details on the technical setup of the observations. 

Observations at 1.1 -- 3.1 GHz are severely affected by radio-frequency interference (RFI), specially from terrestrial and satellite communications. Hence, a careful flagging of RFI was manually applied to the continuum data. This RFI flagging was carried out with the {\scriptsize MIRIAD} package using standard procedures. The flux density scale and the bandpass response were calibrated with the standard flux calibrator PKS 1934--638. The complex-gain calibrators were chosen to be within 7$^\circ$ of each target. The calibrators were PKS 1613--586, MRC 1646--506 and PKS 1710--269. 

We flagged the broadband channels that contained the detected OH emission with the Common Astronomy Software Applications ({\scriptsize CASA}) package, in order to avoid line contamination of the radio continuum. Images were obtained using Brigg's weighting of visibilities with a robust parameter of 0.5 (as defined in {\scriptsize CASA}), and deconvolved using multi-frequency synthesis in task tclean.

Regarding the particular fields containing the objects introduced in Section \ref{sec:candidates}, we noticed the presence of extended galactic background contamination in the data containing the position of IRAS 18019--2216, which significantly worsened the image quality. We thus imaged it again restricting the visibilities to baselines longer than 7 kilolambda.

We also tried in {\scriptsize CASA} phase self-calibration with solution intervals of 10 seconds (the integration time of individual visibilities), which was only successful for IRAS 17494-2645. For the other sources presented in Section \ref{sec:candidates}, we used longer intervals, of 30 seconds. Full widths at half maximum (FWHM) of the synthesized beams in final images vary depending on the individual \textit{uv}  coverage of the observations, but were typically $\simeq$ 3$\arcsec$.5 $\times$ 9$\arcsec$. 

As an additional test, we tried self-calibration and imaging of the final list of OHPNe candidates included in Section \ref{res_cand} using the Astronomical Image Processing System ({\scriptsize AIPS}). No significant difference in signal-to-noise (S/N) ratio was found with respect with the processing with {\scriptsize CASA}.

\begin{table*} 
	\caption{Additional ATCA and VLA data.}
	\label{tab_archive}
	\begin{tabular}{lcllcc}
		\hline
		\hline
		Telescope & Configuration$^a$ & Project Id. & Date & Frequency$^b$ & Bandwidth$^c$\\
		 &  &  & & GHz & MHz\\
		\hline
		 VLA & CnB & AB573  & 1990-Oct-03$^d$ & 4.9 & 25\\
		%		& & & &  1.4 & 25\\
		 ATCA & 6D & C1088  & 2004-Apr-07$^e$ & 4.8 & 128\\
		& & & &  8.6 & 128 \\
		ATCA & 6A & C1450 & 2005-Apr-02$^f$ &  4.8 & 128\\
		& & & &  8.6 & 128 \\
		& & & 2010-Feb-02 & 5.5 & 2000\\
		& & & &  9.0 & 2000\\
		ATCA & 6B & C1610 & 2006-Dec-08 & 1.4 & 128\\ 
		& & & &  2.4 & 128\\
		& &  & 2006-Dec-10 & 4.8 & 128\\
		& & & &  8.6 & 128\\
		ATCA & 6A & C1977  & 2010-Dec-25 & 5.5 &  2000\\
		& & & & 9.0 & 2000\\
		 &  &  & 2010-Dec-28 &  5.5 &  2000\\
		& & & & 9.0 & 2000\\
		VLA & C & 16A-174  & 2016-Jun-09 & 5.8 & 3324\\
		 & & & 2016-Jun-25 & 5.8 & 3324\\
		VLA & D & 17A-197  & 2017-Mar-31 & 5.8 & 3224\\
		ATCA & H168 & C3324  & 2019-Oct-21 & 43.0 & 2000\\
		& & & &  45.0 & 2000\\
		VLA & C & 20A-160  & 2020-Apr-07$^g$  & 3.0 & 2000\\
		\hline
	\end{tabular}
	
	$^a$ Configuration of the antenna array during the observations.\\
	$^b$ Central frequency of the observed band.\\
	$^c$ Total bandwidth.\\
	$^d$ {Results published by \cite*{white05}.}\\
	$^e$ {Results published by \citet{urqu07}.}\\
	$^f$ {Results published by \citet{bains09}.}\\
	$^g$ {Results published by \citet{vlass21}.}\\
\end{table*}

\subsection{Additional VLA and ATCA data}

We searched the archives of ATCA, and \textit{Karl G. Jansky Very Large Array} (VLA)  
for additional radio continuum data at different frequencies, in the fields containing the OHPNe candidates mentioned in section \ref{sec:candidates} within the primary beam of the antennas. We also included data at a wavelength of 7 mm from our own observations of the project C3324 with ATCA. Details of all these additional observations are shown in Table \ref{tab_archive}.

For the published data, we used the flux density values reported in the papers cited in Table \ref{tab_archive}. For the unpublished data, we carried out a complete process of data reduction. 

The unpublished ATCA data were processed, calibrated and imaged following the same procedures mentioned in section \ref{sec:obs_splash}, applying self-calibration when the S/N ratio was high enough. In the particular case of the project code C1977, we combined the visibilities obtained in two different epochs (of observations 3 days apart), to obtain a single image with improved \textit{uv} coverage and sensitivity. For the C3324 observations, performed at 43 and 45 GHz, we tested self-calibration with {\scriptsize CASA} and {\scriptsize AIPS}. It turned out that the final images obtained with {\scriptsize AIPS} procedures had a better S/N ratio, by a factor of about two. Thus, the results presented for those data are the ones obtained with {\scriptsize AIPS} self-calibration and imaging. In particular, we applied amplitude and phase self-calibration with solution intervals of 6 seconds. We note that the definition of the robust parameter of Brigg's weighting in {\scriptsize AIPS} is different from that of {\scriptsize CASA}. We set a robust parameter of zero in {\scriptsize AIPS}, which is equivalent of 0.5 in {\scriptsize CASA}. 

The unpublished archival VLA data from 2016 and 2017 (project codes 16A-174 and 17A-197, respectively) were processed with {\scriptsize CASA}. Calibration was carried out with the VLA pipeline implemented in {\scriptsize CASA}, and further processing was also performed using standard {\scriptsize CASA} procedures. These observations were composed of different fields with phase centers located relatively close to one another. Two of our sources were included within the primary beam of several of these pointings. Thus, for each of our targets, we created mosaic maps with the fields containing them, in order to increase the final S/N ratio at the target position. As in the case of the SPLASH--ATCA data described above, the images containing IRAS 18019-2216 were obtained by using only baselines longer than 7 kilolamda. We attempted self calibration, but we did not obtain any improvement in the image quality of these data.

\section{Results}

\subsection{OHPNe candidates in SPLASH--ATCA}\label{res_cand}

\begin{table*}
	\caption{Confirmed OHPNe and new candidates discovered in SPLASH--ATCA data.}
	\label{tab_candidates}
	\begin{tabular}{llllccc}
		\hline
		\hline
		& Radio continuum emission peak & Maser coordinates$^{b}$ & Transition & $\delta^{c}$ & $\Delta\sigma_{m}$$^{d}$\\
		Source name $^a$  & RA (J2000)\hspace{0.2cm} DEC (J2000) & RA (J2000) \hspace{0.3cm} DEC (J2000) \hspace{0.1cm} & MHz & arcsec & arcsec & Classification$^e$ \\
		\hline
		IRAS 16333--4807 & 16 37 06.603 \hspace{0.1cm} --48 13 42.74 & 16 37 06.581 \hspace{0.1cm} --48 13 42.54 & 1612 & 0.22 & 0.08 & PN \\
		&  & 16 37 06.567 \hspace{0.1cm} --48 13 42.32 & 1667 & 0.42  & 0.03\\
		&  & 16 37 06.588 \hspace{0.1cm} --48 13 42.39 & 1720 & 0.35  & 0.03\\
		&  & 16 37 06.535 \hspace{0.1cm} --48 13 42.35 & 1720 & 0.39  & 0.03\\
		&  & 16 37 06.636 \hspace{0.1cm} --48 13 41.44 & 1720 & 0.30  & 0.03\\
		&  & 16 37 06.628 \hspace{0.1cm} --48 13 42.43 & 1720 & 0.31  & 0.03\\
		&  & 16 37 06.634 \hspace{0.1cm} --48 13 42.46 & 1720 & 0.28  & 0.03\\
		IRAS 16372--4808$^f$ & 16 40 55.812 \hspace{0.1cm} --48 13 59.84 & 16 40 55.824 \hspace{0.1cm} --48 13 59.82 & 1612 & 0.21 & 0.03 & ES\\
		\textbf{OH 341.6811+00.2634} & 16 51 53.637 \hspace{0.1cm} --43 47 16.48 & 16 51 53.641 \hspace{0.1cm} --43 47 16.47 & 1612 & 0.11 & 0.19 & U\\
		JaSt 23 & 17 40 23.068 \hspace{0.1cm} --27 49 12.13 & 17 40 23.057 \hspace{0.1cm} --27 49 12.16 & 1612 & 0.04 & 0.03 & PN\\
		&  & 17 40 23.064 \hspace{0.1cm} --27 49 12.18 & 1612 & 0.09  & 0.03\\
		&  & 17 40 23.056 \hspace{0.1cm} --27 49 12.87 & 1665 & 0.86 & 0.03\\
		IRAS 17375--2759 & 17 40 38.575 \hspace{0.1cm} --28 01 04.91 & 17 40 38.562 \hspace{0.1cm} --28 01 02.28 & 1612 & 2.64 & 0.03  & ES$^g$ \\
		&  & 17 40 38.591 \hspace{0.1cm} --28 01 02.39 & 1667 & 2.54  & 0.03 \\
		IRAS 17393--2727 & 17 42 33.139 \hspace{0.1cm} --27 28 25.34 & 17 42 33.123 \hspace{0.1cm} --27 28 25.05 & 1612 & 0.31 & 0.03 & PN\\
		&  & 17 42 33.124 \hspace{0.1cm} --27 28 25.03 & 1612 & 0.33 & 0.03 \\
		&  & 17 42 33.120 \hspace{0.1cm} --27 28 25.02 & 1612 & 0.34  & 0.03 \\
		&  & 17 42 33.125 \hspace{0.1cm} --27 28 25.17 & 1612 & 0.18  & 0.03 \\
		&  & 17 42 33.117 \hspace{0.1cm} --27 28 25.13 & 1612 & 0.21  & 0.03 \\
		&  & 17 42 33.110 \hspace{0.1cm} --27 28 25.11 & 1612 & 0.26  & 0.03 \\
		&  & 17 42 33.121 \hspace{0.1cm} --27 28 25.22 & 1665 & 0.14  & 0.03 \\
		&  & 17 42 33.100 \hspace{0.1cm} --27 28 25.41 & 1667 & 0.11  & 0.03 \\
		&  & 17 42 33.115 \hspace{0.1cm} --27 28 25.03 & 1667 & 0.33  & 0.03 \\
		&  & 17 42 33.119 \hspace{0.1cm} --27 28 24.93 & 1667 & 0.43  & 0.03 \\
		&  & 17 42 33.100 \hspace{0.1cm} --27 28 25.06 & 1667 & 0.31  & 0.03 \\
		\textbf{IRAS 17494--2645} & 17 52 30.817 \hspace{0.1cm} --26 46 17.68 & 17 52 30.815 \hspace{0.11cm} --26 46 16.68 & 1612 & 1.00 & 1.00 & U\\
		\textbf{IRAS 18019--2216} & 18 04 57.391 \hspace{0.1cm} --22 15 50.97 & 18 04 57.387 \hspace{0.1cm} --22 15 51.28 & 1612 & 0.31 & 0.25 & U\\
		&  & 18 04 57.391 \hspace{0.1cm} --22 15 50.58 & 1612 & 0.39  & 0.03 \\
		&  & 18 04 57.413 \hspace{0.1cm} --22 15 49.41 & 1612 & 1.19  & 0.03 \\
		\hline
	\end{tabular}
	
	$^a$ Source names in bold face indicate those for which the association between OH maser and radio continuum emission is reported here for the first time.\\
	$^b$ Coordinates reported in \cite{qia16b,qia18,qia20}.\\
	$^c$ Relative distance between the radio continuum emission peak and the OH maser spot.\\
	$^d$ Relative positional accuracy between maser and continuum at 2.1 GHz.\\
	$^e$ Classification in \cite{qia16b,qia18,qia20}. ES, PN and U represent evolved star, planetary nebula and object of unknown nature, respectively.\\
	$^f$ The association between radio continuum and OH maser emission was reported by \cite{bains09}, although these authors did not consider it to be OHPN. We discuss this source in our paper as new OHPN candidate.\\
	$^g$ The nature of this source as a PN will be discussed in Cala et al. (in preparation).\\
\end{table*}

\label{sec:candidates}
\begin{figure*}
	\centering
	\includegraphics*[width=0.42\textwidth]{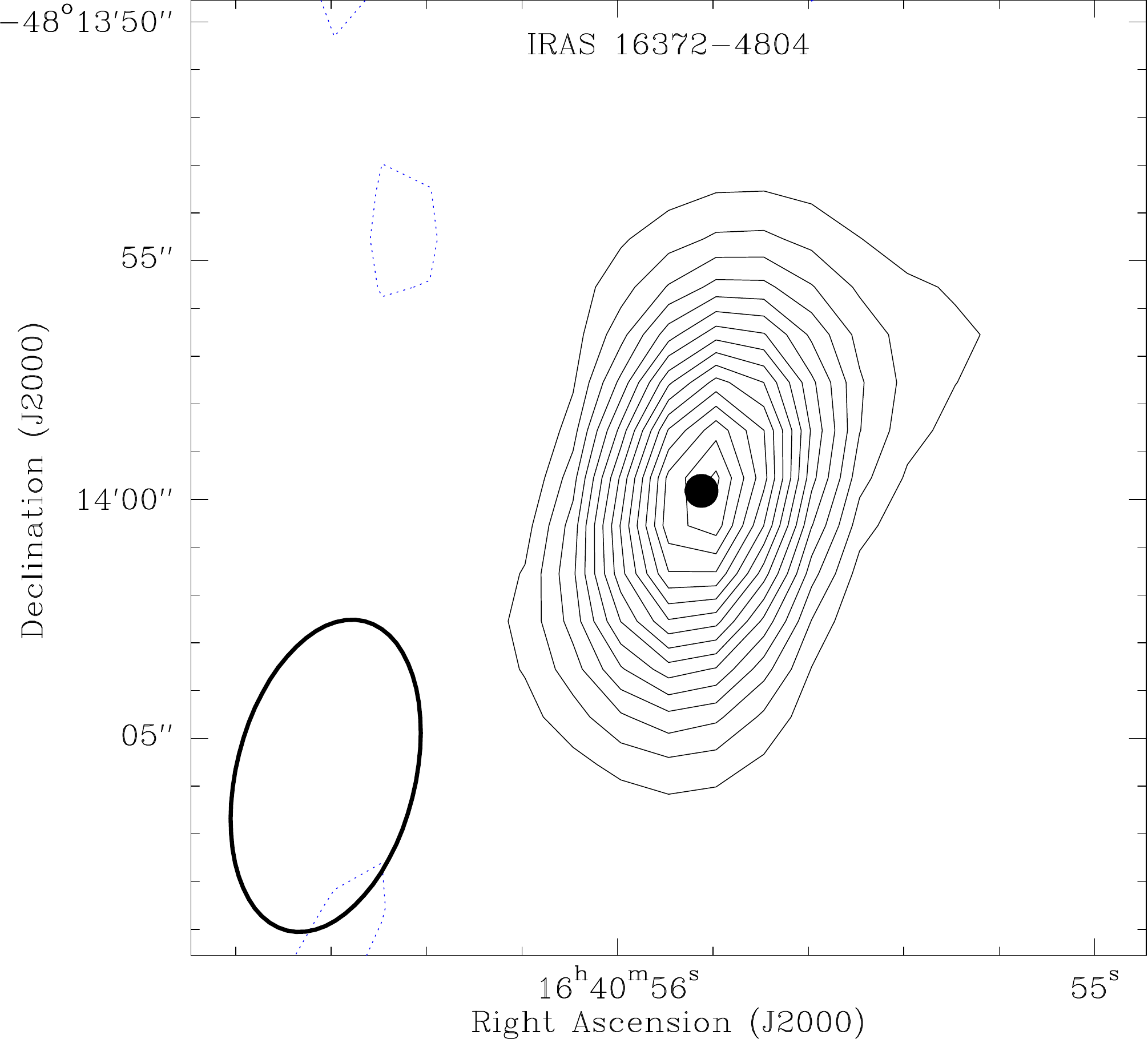}
	\includegraphics*[width=0.42\textwidth]{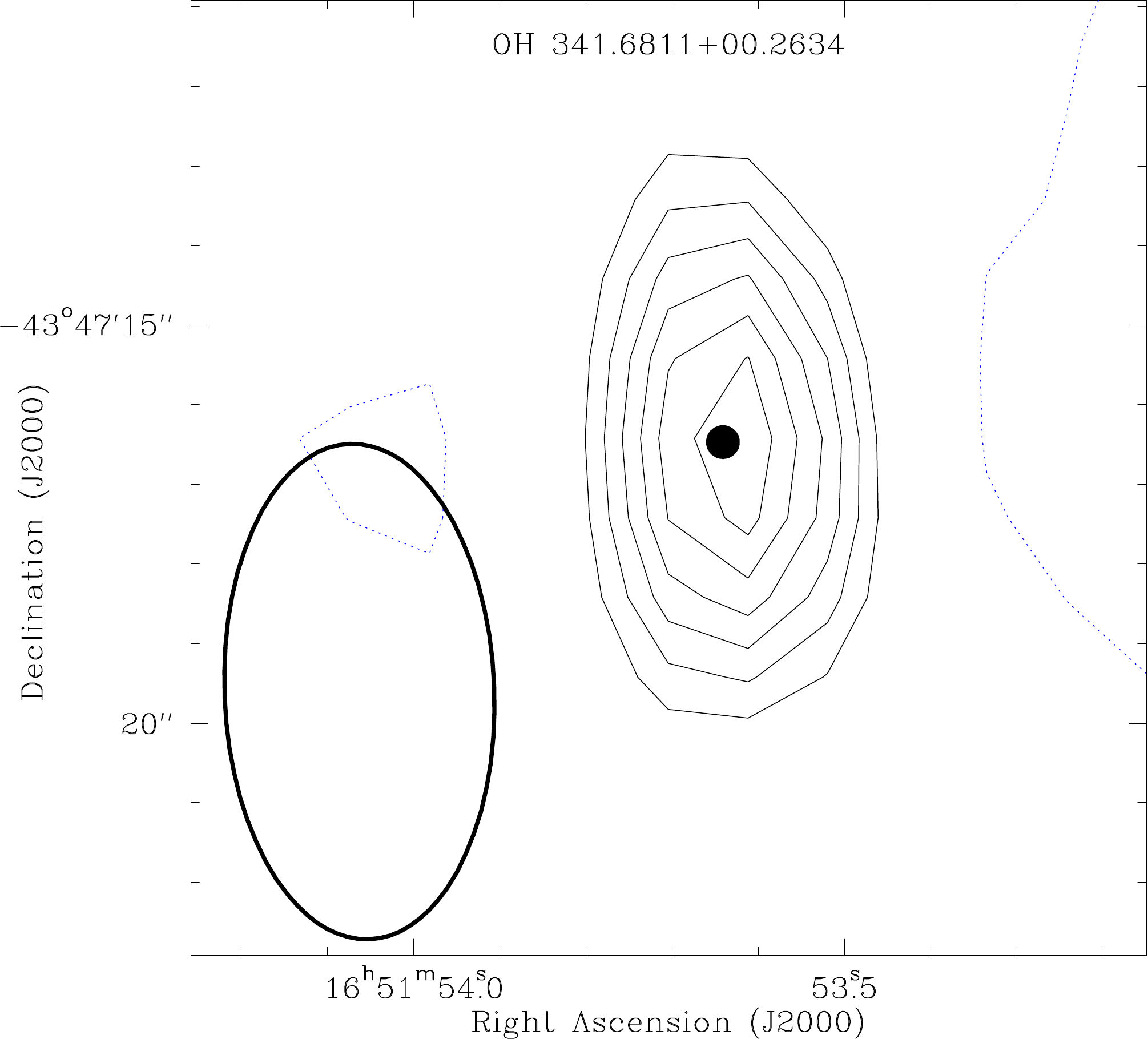}
	\includegraphics*[width=0.42\textwidth]{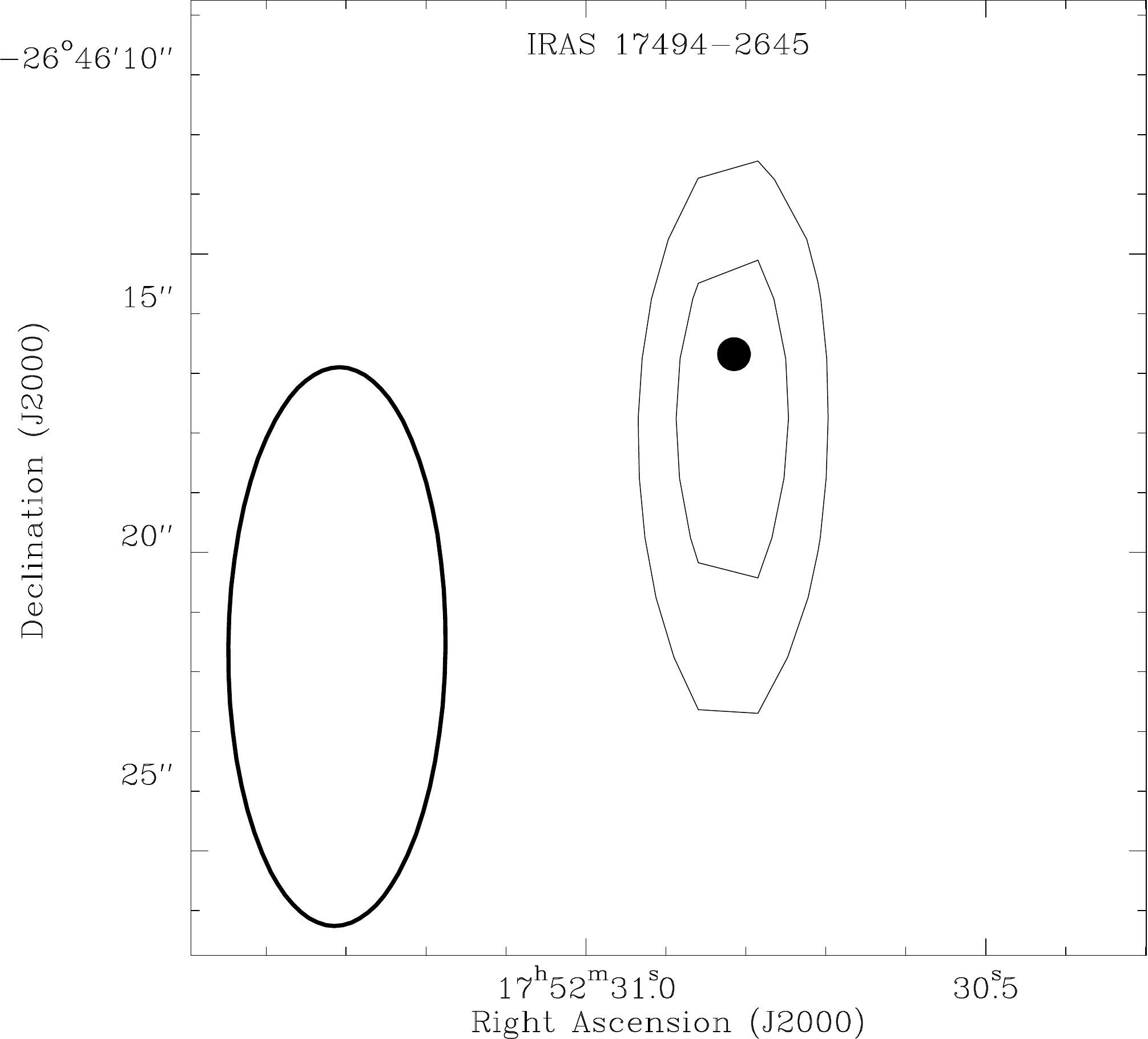}
	\includegraphics*[width=0.42\textwidth]{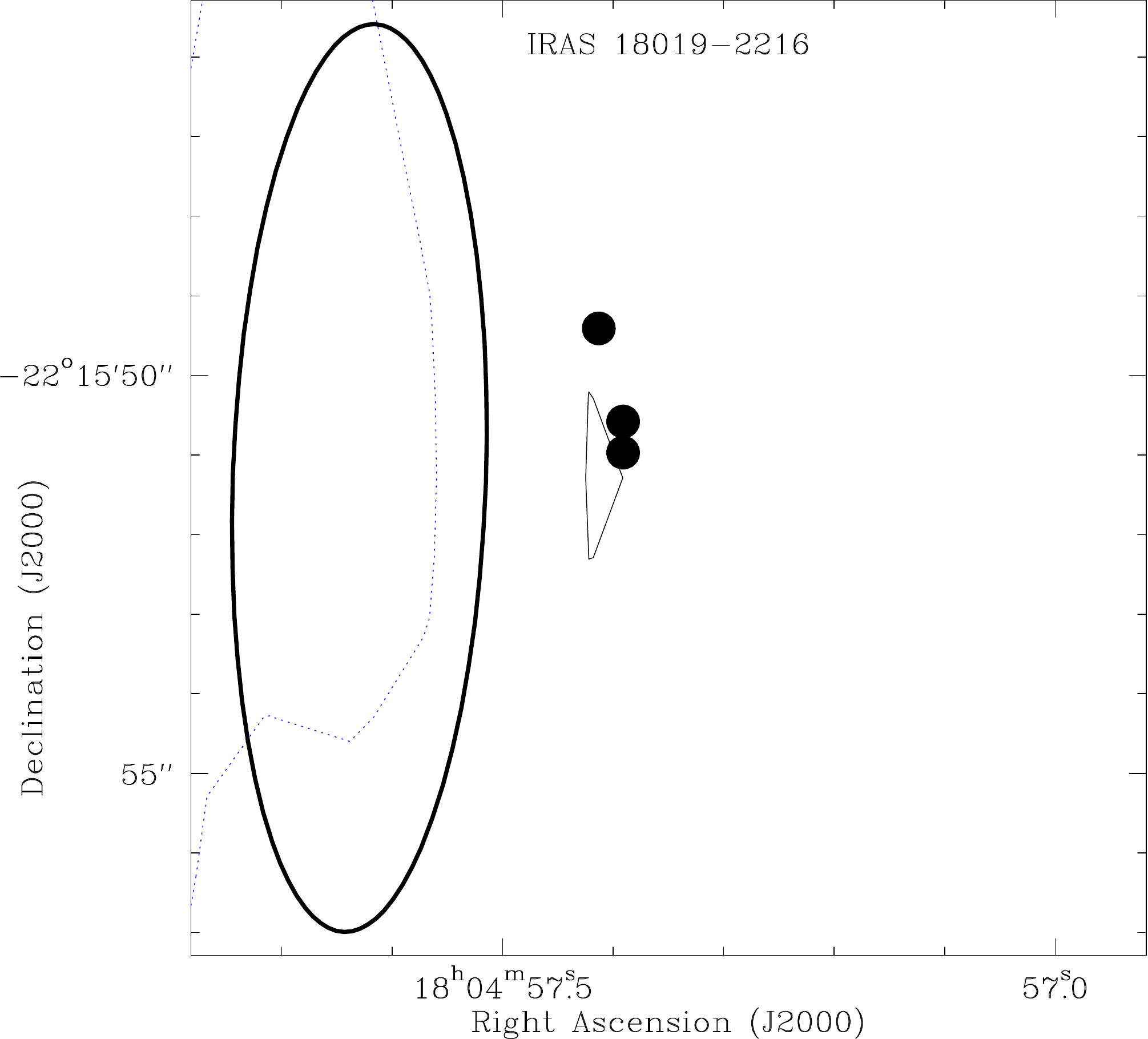}
	\caption{Contour maps of the radio continuum emission at 2.1 GHz of the new OHPNe candidates (Table \ref{tab_candidates}). The black filled circles mark the location of the spectral OH maser features at 1612 MHz. The thick-lined ellipse represents the FWHM of the synthesized beam of each continuum image. Solid black and dotted blue contours represent positive an negative intensity values, respectively. \textit{Top left}. IRAS 16372--4808: the first contour level and the increment step are 2 times $\sigma$ (where $\sigma= 0.24$ mJy beam$^{-1}$ is the rms of the map). The beam size is (6$\arcsec$.68 $\times$ 3$\arcsec$.75, p.a. = --14$\degree$). \textit{Top right}. OH 341.6811+00.2634: the first contour level is $2\times\sigma$ and the increment step is $1\times\sigma$ ($\sigma = 0.31$ mJy beam$^{-1}$). The beam size is (6$\arcsec$.23 $\times$ 3$\arcsec$.39, p.a. = 2$\degree$). \textit{Bottom left}. IRAS 17494--2645: the contour levels are 2 and $3\times\sigma$ ($\sigma = 0.31$ mJy beam$^{-1}$). The beam size is (9$\arcsec$.36 $\times$ 3$\arcsec$.64, p.a. = 0$\degree$). \textit{Bottom right}. IRAS 18019--2216: the contour levels are --2 and 2$\times\sigma$ ($\sigma=0.20$ mJy beam$^{-1}$). The beam size is (11$\arcsec$.41 $\times$ 3$\arcsec$.18, p.a. = --2$\degree$). See also Fig. \ref{fig:corr_vla}. }
	\label{fig:corr}
\end{figure*}

We searched in the processed continuum images the catalogued positions of the OH maser-emitters sources of \cite{qia16b, qia18, qia20}. We kept only those where the coincidence between the OH maser relative to the continuum source were smaller than the FWHM synthesized beam ($\theta$ $\simeq$ 3$\arcsec$.5 $\times$ 9$\arcsec$) of the continuum. Afterwards, we discarded all objects classified as SFRs or supernova remnants in \cite{qia16b, qia18, qia20}, and kept only with those identified as evolved stars or objects of unknown nature that could be evolved stars. We also rejected two cases (G359.567+1.147 and G008.483+0.176) of relatively close spatial match between maser and continuum emission (within $\simeq 5''$), even if the maser might be associated with an evolved star, but a literature search showed that the continuum emission traces an extragalactic object. We also detect IRAS 18043--2116, a WF post--AGB star \citep{wal09}, whose radio continuum emission has been studied by \citet{per17}. The remaining sources with a positive match, are reported in Table \ref{tab_candidates}. 

This match criterion of OH masers being within one synthesized beam of the continuum emission seems adequate, since interferometric observations consistently show at least one OH spectral component and OH transition very close to the peak of the radio continuum emission in OHPNe \citep[$\la 1''$; ][]{usc12,gom16}. Moreover, OHPNe are angularly small in images (at optical, infrared or radio wavelengths), with sizes extending $\la 3''$ from the nebular center \citep[e.g.,][]{mir01,deG04,usc14}. Thus, even if OH emission was to be excited at the tip of the lobes, our match criterion would associate continuum and maser emission. In any case, the largest distance between OH and continuum emission in Table \ref{tab_candidates} is $\sim$1$\arcsec$, whereas the shortest distance for sources that did not fulfil our criterion is $\sim$15$\arcsec$. Therefore, we believe that our distance criterion did not reject any true association, whereas false positives are highly unlikely.

Table \ref{tab_candidates} gives the parameters of the continuum emission peaks at 2.1 GHz for these sources, as well as the coordinates of the OH maser components associated with the continuum sources. The distances of the maser components to the continuum sources ($\delta$) are also derived. Since the maser and continuum observations were simultaneous, they share the same interferometric fringe phase calibration. Therefore, the 1-$\sigma$ error of the relative position between maser and continuum emission is $\Delta\sigma{_m}\sim \theta/(2\times \rm{S/N})$ \citep{reid88}, where $\theta$ is the apparent source size (corresponding to the FWHM of the beam for unresolved sources), and S/N is the signal-to-noise ratio of the emission. The absolute astrometric accuracy in SPLASH--ATCA is $< 1 \arcsec$ \citep{qia16b}.

\begin{figure}
	\centering
	\includegraphics*[width=0.45\textwidth]{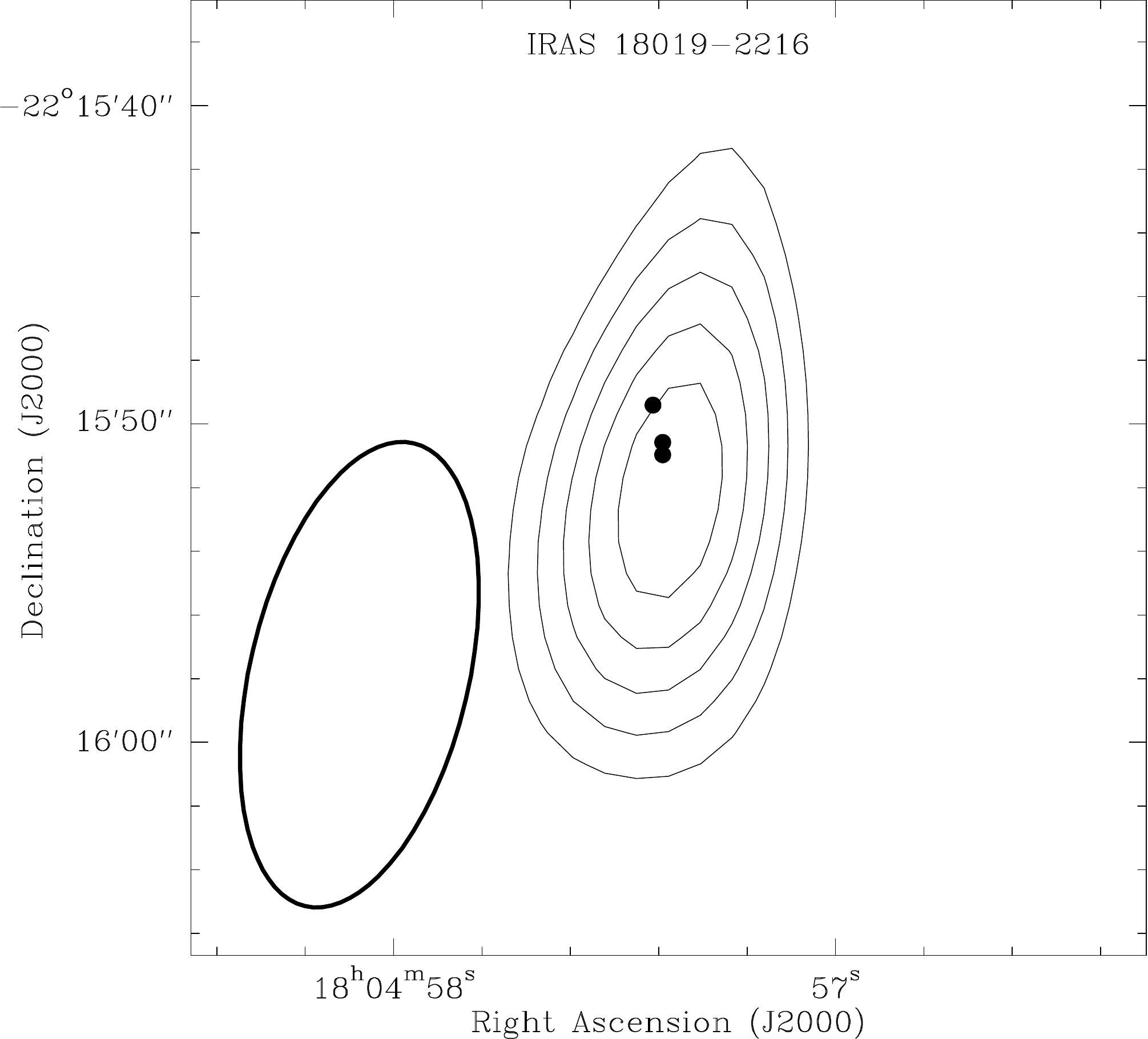}
	\caption{Contour map of the continuum emission at 5 GHz in IRAS 18019--2216, taken on 2017 March 31. The black circles mark the location of the spectral OH maser features at 1612 MHz. The first contour level is $2\times \sigma$, and the increment step is $1\times \sigma$ ($\sigma= 90$ $\mu$Jy beam$^{-1}$. The thick-lined ellipse represents the FWHM of the synthesized beam of the continuum image (14$\arcsec$.93 $\times$ 6$\arcsec$.87, p.a. = --13$\degree$). }
	\label{fig:corr_vla}
\end{figure}

In total, over the area of 176 deg$^2$ covered by SPLASH--ATCA, we found a match between OH maser and radio continuum emission in eight objects that are likely to be OHPNe. Three of these objects are already known OH emitters that are bona-fide PNe \citep[IRAS 17393--2727, JaSt 23, and IRAS 16333--4807;][]{usc12,qia16a}. The association of OH maser and continuum emission in IRAS 17375--2759 was confirmed by \citet{usc12}, but these authors classified it as an OHPN candidate. The nature of this source as a PN will be discussed in a forthcoming paper (Cala et al. in preparation). Three more objects of unknown nature are reported here for the first time: OH 341.6811+00.2634, IRAS 17194--2645, and IRAS 18019--2216. Finally, for IRAS 16372--4808 the association between OH maser and radio continuum emission was reported by \citet{usc12}, but it was not considered to be an OHPN, based on its previous classification as a post--AGB star by \citet{bains09}. However, as discussed below, there is not enough evidence to discard the nature of IRAS 16372--4808 as PN. Considering this re-evaluation of the source, we included it in the remaining of this paper as a new OHPN candidate.

Radio continuum images at 2.1 GHz of these 4 new OHPN candidates using SPLASH--ATCA data are shown in Fig. \ref{fig:corr}. In the case of IRAS 18019--2216, this emission was only at a 2-sigma level, although coincident with the position of the maser components. The archival VLA observations at 5 GHz have confirmed the presence of radio continuum emission (Fig. \ref{fig:corr_vla}).

\subsection{Radio continuum emission and spectral indices} \label{sec:spec_ind}

\begin{figure*}
	\begin{centering}
		\includegraphics*[width=0.49\textwidth]{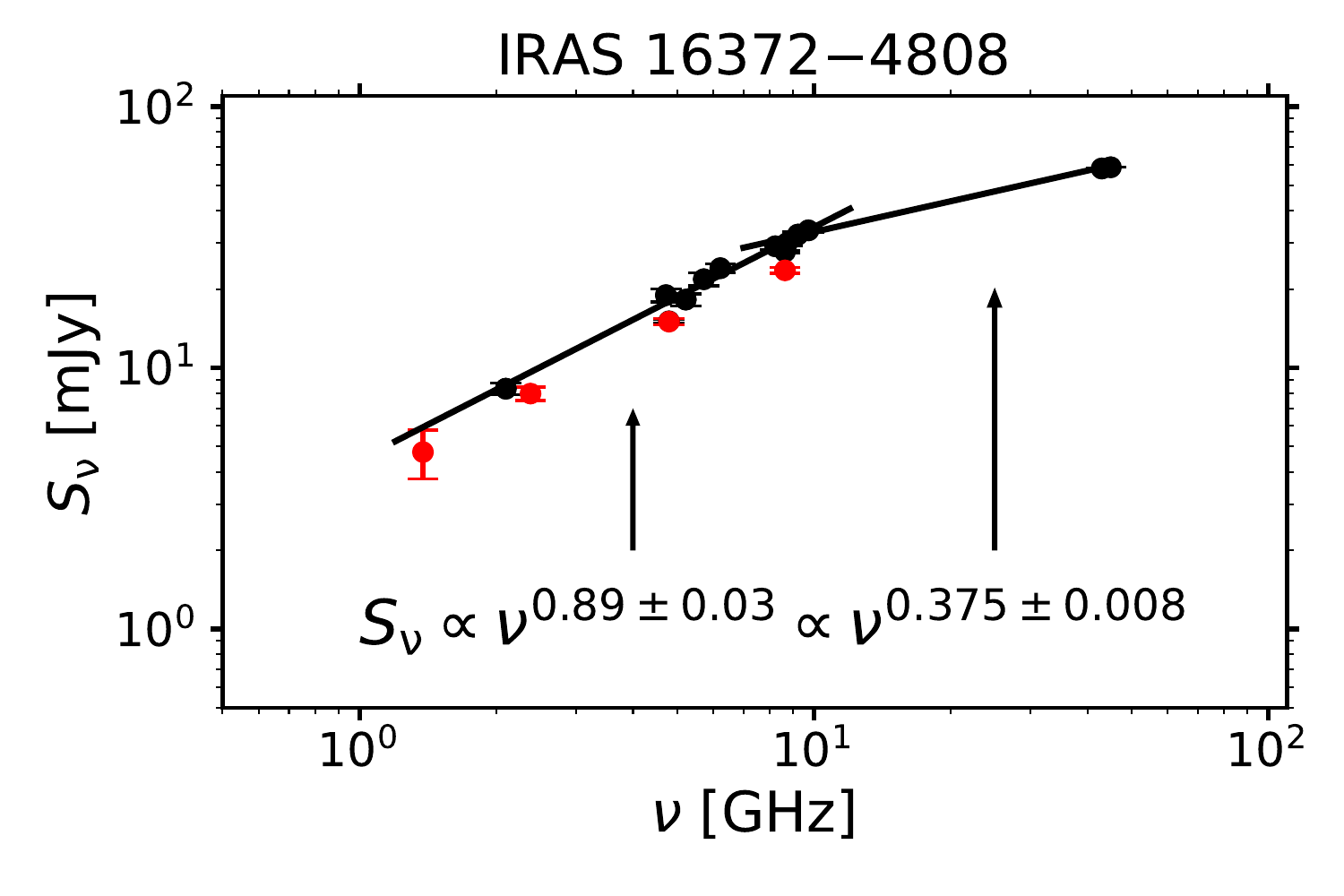}
		\includegraphics*[width=0.49\textwidth]{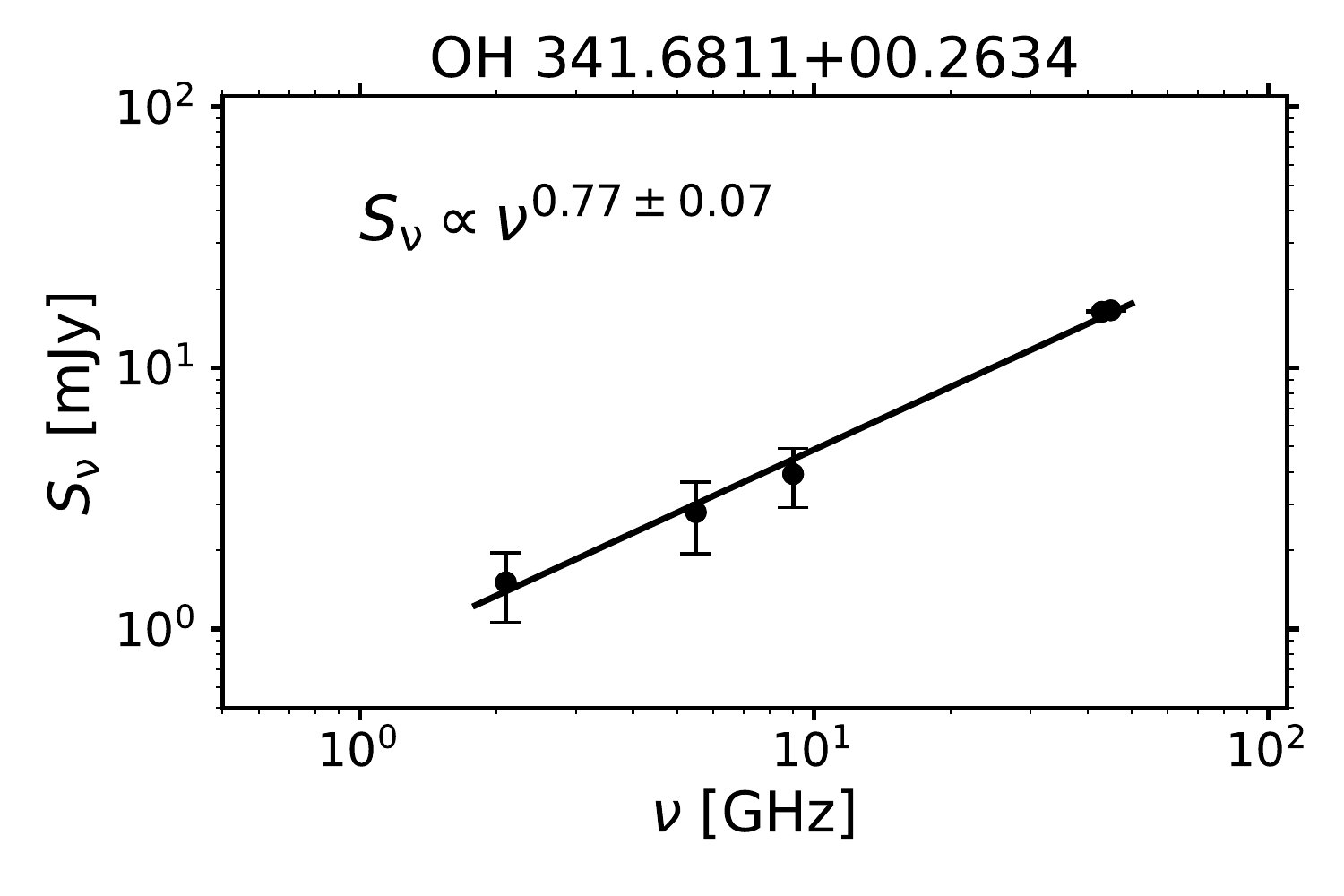}
		\includegraphics*[width=0.49\textwidth]{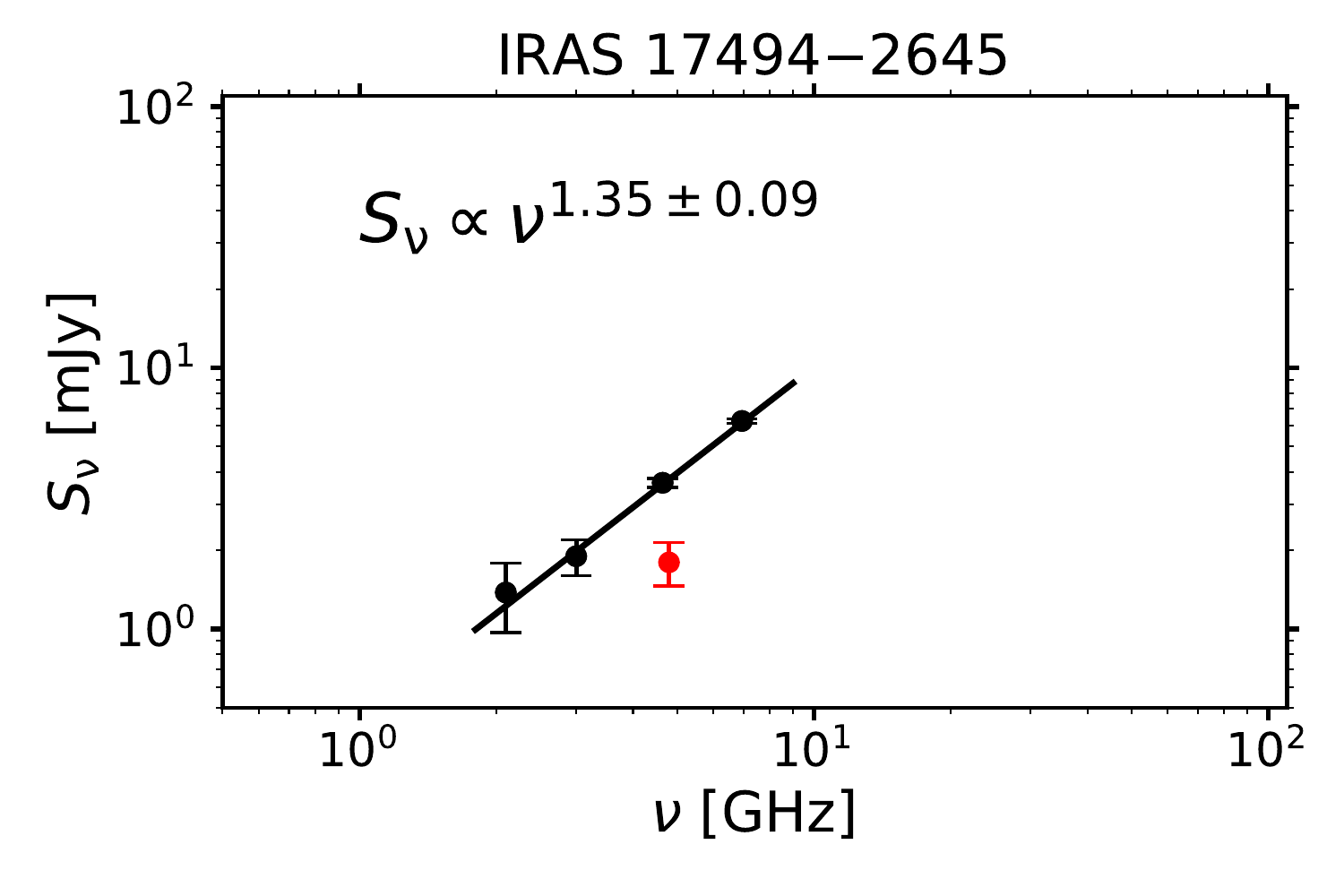}
		\includegraphics*[width=0.49\textwidth]{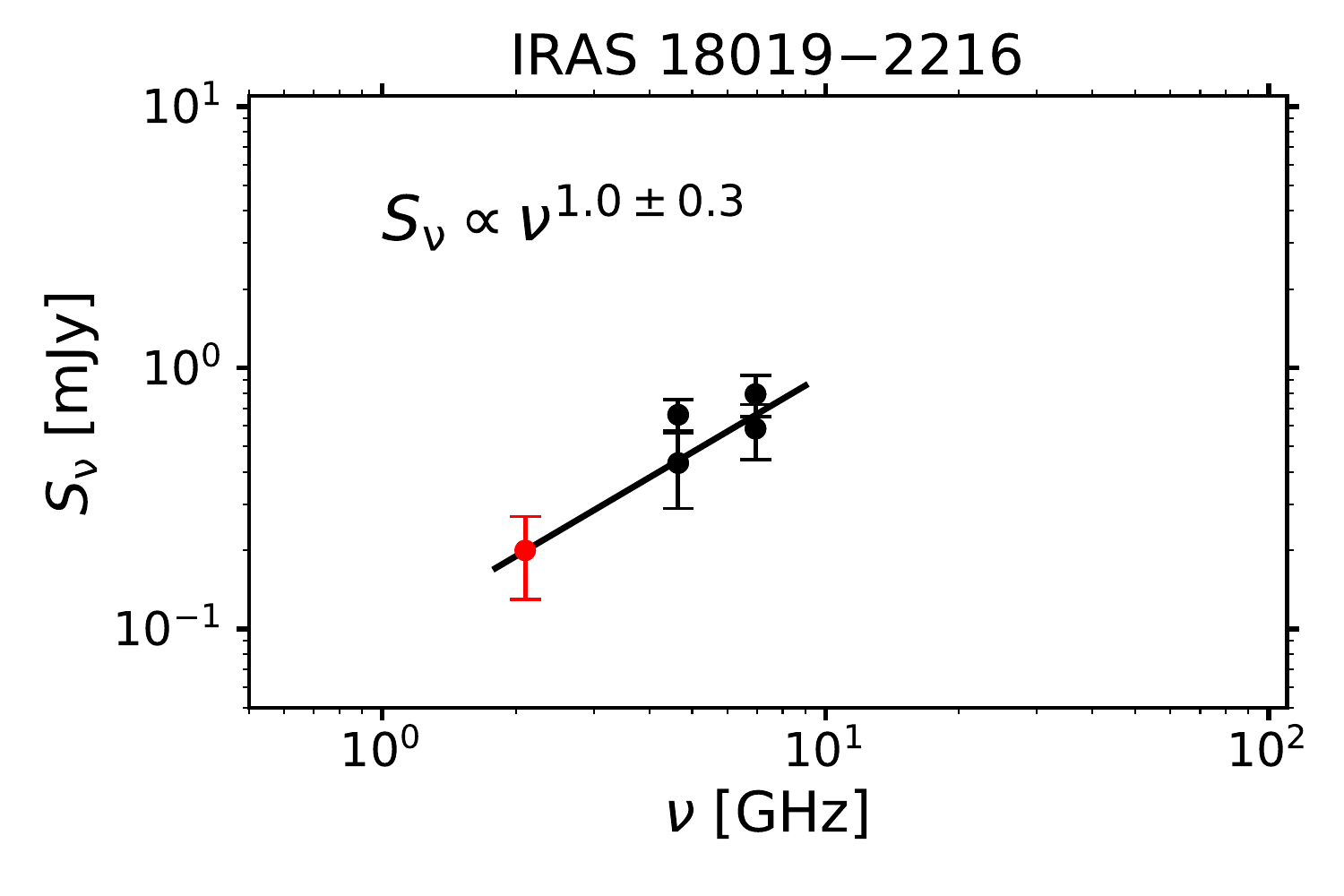}
		\caption{Radio continuum spectra of the new OHPNe candidates. Black and red circles represent the flux densities and its errors at different epochs (see below and Table \ref{tab_fluxes}). In some data points, the errors are smaller than the symbol sizes. The solid lines are least-square linear fits to $\log(S_\nu)$ vs $\log(\nu)$, whose slope gives the estimated spectral index. These fits were obtained using the data marked as black circles (see below). Notes: \textit{IRAS 16372--4808:} Spectral indices shown in the figure were obtained with data taken on or after 2010 (black points), resulting in $\alpha$ = 0.89 $\pm$ 0.03 between 2.1 -- 9.7 GHz, and $\alpha$ = 0.39 $\pm$ 0.04 between 8.7 -- 45 GHz. A comparison with data taken on or before 2006 (red points)  shows an increase of flux density with time, while maintaining a similar spectral index of 0.082 $\pm$ 0.04. \textit{IRAS 17494--2645:} The red circle represents the flux density at 5 GHz in 1990, which seems to have increased by a factor of 2.0 $\pm$ 0.45 in 26 yr. \textit{IRAS 18019--2216:} The red point represents the flux density detected with an S/N $\simeq$ 2 at 2.1 GHz in this paper. The spectral index was obtained using at 4.65 and 6.95 GHz the arithmetic mean (weighted by the inverse of noise squared) of data at the different epochs. A value of $\alpha$ = 0.59 $\pm$ 0.75 is obtained if the value of flux density at 2.1 GHz is ignored in the computation.} 
	
	\label{fig:cont_spec}
	\end{centering}
\end{figure*}

We have explored the flux density (S$_{\nu}$) dependence on frequency ($\nu$) of the radio continuum emission. In data with high S/N, the observed band of individual observations was separated on smaller frequency subbands ($\simeq 500$ MHz). The dependence of flux density on frequency is presented in Table \ref{tab_fluxes} and Fig. \ref{fig:cont_spec}. The resulting spectral indices, $\alpha$ (defined as $S_{\rm \nu} \propto \nu^{\alpha}$) are also presented in the individual panels of that figure.

\begin{table*}
	\caption{Radio continuum flux density of the OHPNe candidates.}
	\label{tab_fluxes}
	\begin{tabular}{llcccr}
		\hline
		\hline
				&	& Frequency & $S_{\nu}$ & rms \\
		No. & OHPNe candidate & GHz & mJy & mJy beam$^{-1}$ & Date\\
			\hline
		1 & IRAS 16372--4808 & 1.38 & 4.8 $\pm$ 1.0 & 0.4 & 2006-Dec-08\\
					&	& 2.10 & 8.3 $\pm$ 0.4 & 0.24 & 2013-Oct-25\\
					&	& 2.38 & 10.0 $\pm$ 0.5 & 0.21 & 2006-Dec-08\\
					&	& 4.73 & 19.0 $\pm$ 1.1 & 0.5 & 2010-Feb-02\\
					&	& 4.80 & 15.10 $\pm$ 0.5 & 0.3 & 2004-Apr-07$^{a}$\\
					&	&  & 15.17 $\pm$ 0.20 & 0.10 & 2005-Apr-02$^{b}$\\
					&	&  & 15.03 $\pm$ 0.4 & 0.18 & 2006-Dec-10\\
					&	& 5.23 & 18.25 $\pm$ 1.0 & 0.5 & 2010-Feb-02\\
					&	& 5.73 & 21.8 $\pm$ 1.25 & 0.4 & 2010-Feb-02\\
					&	& 6.23 & 24.1 $\pm$ 1.0 & 0.4 & 2010-Feb-02\\
					&	& 8.23 & 29.20 $\pm$ 0.8 & 0.22 & 2010-Feb-02\\
					&	& 8.64 & 27.8 $\pm$ 0.6 & 0.3 & 2004-Apr-07$^{a}$\\ 
					&	&  & 24.6 $\pm$ 0.5 & 0.25 & 2005-Apr-02$^{b}$\\
					&	&  & 23.6 $\pm$ 0.6 & 0.21 & 2006-Dec-10\\
					&	& 8.73 & 30.1 $\pm$ 0.7 & 0.21 & 2010-Feb-02\\
					&	& 9.23 & 32.4 $\pm$ 0.7 & 0.23 & 2010-Feb-02\\
					&	& 9.73 & 33.6 $\pm$ 0.7 & 0.23 & 2010-Feb-02\\
					&	& 43.0 & 58.0 $\pm$ 0.1 & 0.03 & 2019-Oct-21\\
					&	& 45.0 & 58.6 $\pm$ 0.034 & 0.01 & 2019-Oct-21\\
		2 & OH 341.6811+00.2634 & 2.10 & 1.5 $\pm$ 0.5 & 0.3 & 2013-Oct-28\\
					&	  & 5.50 & 2.8 $\pm$ 0.9 & 0.6 & 2010-Dec-28\\
					&	  & 9.00 & 3.9 $\pm$ 1.0 & 0.7 & 2010-Dec-28\\
					&	  & 43.0 & 16.4 $\pm$ 0.1 & 0.03 & 2019-Oct-21\\
					&	  & 45.0 & 16.6 $\pm$ 0.13 & 0.1 & 2019-Oct-21\\
		3 & IRAS 17494--2645   & 2.10 & 1.4 $\pm$ 0.4 & 0.3 & 2006-Feb-26\\
		             &      & 3.0$^{c}$ & 1.9 $\pm$ 0.3 & 0.01 & 2020-Apr-07\\
					&	  & 4.65 & 3.6 $\pm$ 0.14 & 0.1 & 2016-Jun-09\\
					&	  & 4.85$^{d}$ & 1.8 $\pm$ 0.34 & 0.3 & 1990-Oct-03\\
					&	  & 6.95 & 6.2 $\pm$ 0.13 & 0.1 & 2016-Jun-09\\
		4 & IRAS 18019--2216   & 2.10 & 0.2 $\pm$ 0.07$^{e}$ & 0.1  & 2016-Mar-01\\
					&	  & 4.65 & 0.7 $\pm$ 0.1 & 0.1 & 2016-Jun-25\\
					&	  & & 0.4 $\pm$ 0.14 & 0.1 & 2017-Mar-31 \\
  					&	  & 6.95 & 0.8 $\pm$ 0.14 & 0.1 & 2016-Jun-25\\
  					&	  &  & 0.6 $\pm$ 0.14 & 0.1 & 2017-Mar-31\\
		\hline
	\end{tabular}

		$^{a}$ Reported in \cite{urqu07}. \\
		$^{b}$ Reported in \cite{bains09}. \\
		$^{c}$ Reported in \cite{vlass21}.\\
		$^{d}$ Reported in \cite{white05}.\\
		$^{e}$ Detected with S/N $\simeq$ 2.

\end{table*}

The spectral indices of the radio continuum provide useful information on the nature of these sources. For all four OHPNe candidates of SPLASH, they are consistent with the expected values for thermal (free-free) radiation from photoionized gas, which ranges between $\alpha = +2$ (optically thick regime) at low frequencies and $\alpha =-0.1$ (optically thin regime) at high frequencies. The values we obtained suggest the presence of an ionized region that is partially optically thick at radio wavelengths. 

For ionized regions with nearly constant electron density, one would expect a relatively abrupt transition between the optically thick emission at low frequencies and the optically thin regime at high ones. This is the common trend observed in PNe \citep[see, e.g.,][]{aa91}. This transition between opacity regimes determines a turnover frequency that is directly related to the emission measure of the ionized gas ($EM=\int n_e^2 dl$, where $n_e$ is the electron density profile, and $l$ is the linear distance along the line of sight). Thus, a higher $n_e$ would result in a higher turnover frequency. In our candidates, the radio spectrum does not flatten to $\alpha =-0.1$, at the frequencies sampled (up to $\sim$ 2 -- 45 GHz). Such high turnover frequencies are considered to be a sign of youth in PNe \citep{kwo81b}.

We also note that the spectral indices are relatively constant over a wide range of frequencies. This behaviour is also consistent with these objects being nascent PNe.
At the very first stages of photoionization of the previously expelled CSE, the electron density of the ionized gas should be consistent with that of the envelope ($n_{\rm e} \propto r^{-2}$, where $r$ is the distance from the central star). An ionized region with that radial dependence would be partially optically thick over a wide range of frequencies, with spectral index $\alpha \simeq 0.6$ \citep{ol75}. 
The radio spectrum of the new OHPNe candidates shown in Fig \ref{fig:cont_spec}  are, in some cases, relatively close to that value of $\alpha = 0.6$, specially for OH 341.6811+00.2634. The data at 43--45 GHz for IRAS 16372--4808 suggest that the source is becoming optically thinner at high frequencies, but a better frequency sampling above 10 GHz would be needed to accurately determine the spectral index in that range.
These spectral indices $\sim 0.6$, are also found in confirmed OHPNe \citep[e.g., K 3--35 and IRAS 17347-3139;][]{aa93, gom05, taf09}. Assuming a power-law distribution of ionized gas density with spherical symmetry \citep[$n_{\rm e} \propto r ^{-q}$;][]{ol75}, values for $q$ =  2.2, 2.4 and 2.6 are obtained for OH 341.6811+00.2634, IRAS 16372--4808, and IRAS 18019--2216, using the obtained spectral indices $\alpha \simeq $ 0.77, 0.89, and 1.0 respectively. 
For the case of IRAS 17494--2645, the spectral index of 1.35 would imply, in principle, a steep power-law index $q=3.4$. However, the emission of this source is supposed to be in the optically thick regime, and the spectral index may not reflect such a radial dependence on electron density. This could happen if the electron density is higher than those in the other candidates. We speculate that observations at $\nu > 10$ GHz may derive a flatter spectral index, closer to 0.6. 

Another interesting property is time variation of the radio continuum emission. We have consistently found an increase of flux density, both in the four new OHPNe candidates discussed here, as well as several of the already confirmed OHPNe in Table \ref{tab_candidates}. Regarding the known OHPNe whose continuum emission we detected in SPLASH--ATCA, the flux density at 2.1 GHz of JaSt 23 has increased, in around 3.5 yr, by a factor of 1.49 $\pm$ 0.26, with respect to the value reported by \cite{gom16}.

For our new OHPNe candidates, in IRAS 16372--4808 the flux densities at 4.8 and 8.4 GHz increased by a factor of 1.26 $\pm$ 0.11 and 1.26 $\pm$ 0.06 in around 6 and 5 yr, respectively (see Table \ref{tab_fluxes}). \cite{white05} reported for IRAS 17494--2645 an $S_{\nu}$ = 1.8 $\pm$ 0.3 at 5 GHz. We obtained $S_{\nu}$ = 3.6 $\pm$ 0.14 mJy and $S_{\nu}$ = 6.3 $\pm$ 0.14 mJy at 4.65 and 6.95 GHz, respectively (see Table \ref{tab_fluxes}). Hence, the flux density of this source has increased by a factor of 2.0 $\pm$ 0.45 in around 26 yr. These variations are further discussed in Section \ref{sec:description_candi}. 

The increase in flux density of radio continuum emission in nascent PNe could be interpreted as expansion of the ionization front. In this case, the flux density would be proportional to $R^2$, where $R$ is the radius of the ionized region. This was the interpretation given by \citet{gom05} and \citet{taf09} for the increasing emission in the OHPN IRAS 17347--3139. This source also fell within the sky coverage of SPLASH, although we did not detect its OH maser emission, which was weaker than the detection limit of the survey. We did detect, however, its continuum emission, and confirmed its increasing flux density, with $S_{\rm \nu}$ = 54.5 $\pm$ 4.4 mJy. This means that, with respect to the flux density expected at 2.1 GHz in \cite{taf09}, it has increased by a factor of 1.74 $\pm$ 0.23 in around 11 yr. This increase is roughly consistent with the factor of $1.31\pm 0.05$ at 9.4 GHz in 13 yr found by \citep{gom05}.

It is obvious that, considering this variability, spectral indices should be taken with care. A more accurate determination of spectral indices would require observations that are simultaneous or at least taken over a very short time interval. Despite this intrinsic uncertainty in obtaining numerical values, the radio spectra in Fig. \ref{fig:cont_spec} show robust trends for these objects.

\subsection{Optical-Infrared spectral energy distribution (SED)}\label{sec:red}

\begin{figure*}
	\includegraphics*[width=0.49\textwidth]{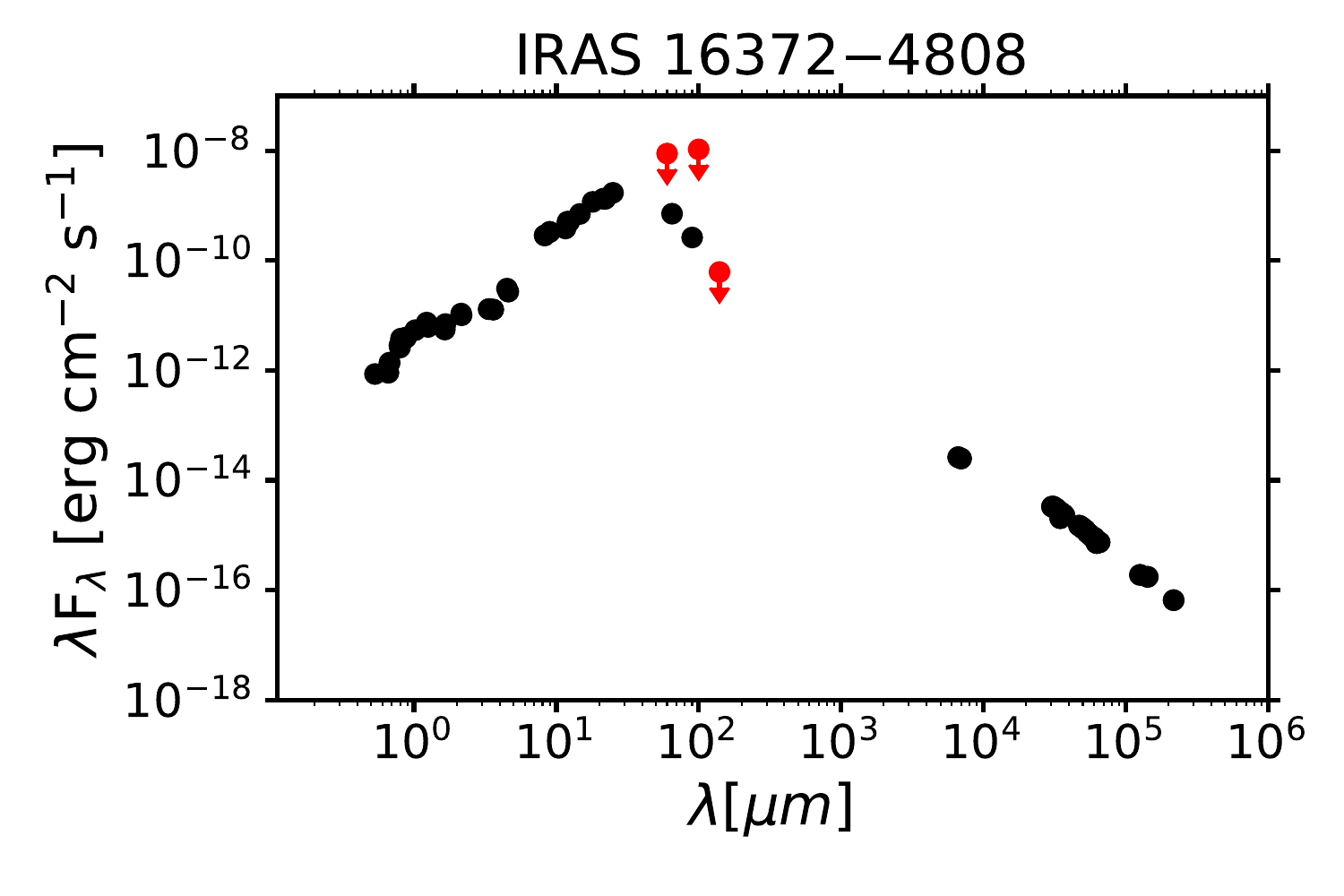}
	\includegraphics*[width=0.49\textwidth]{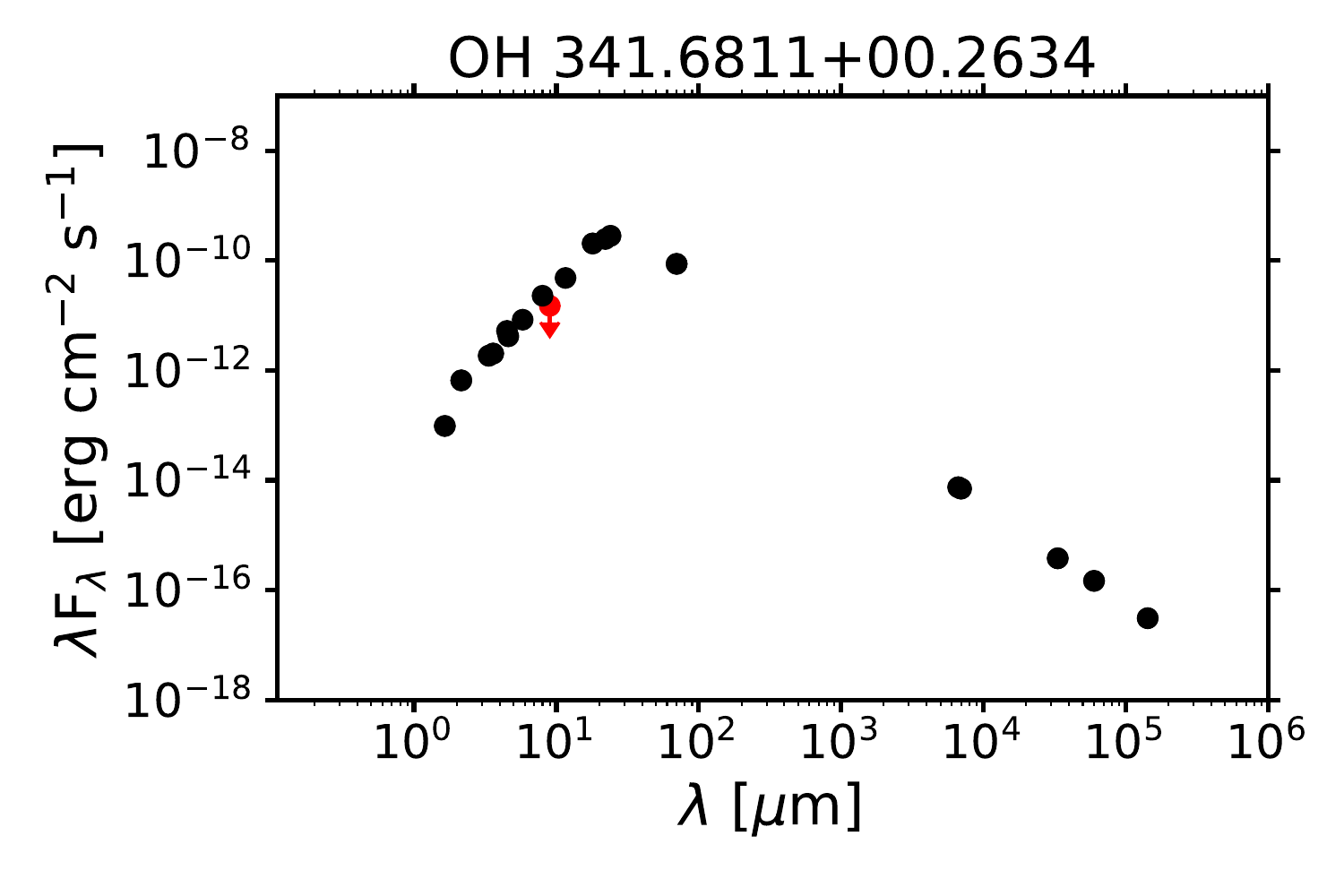}
	\includegraphics*[width=0.49\textwidth]{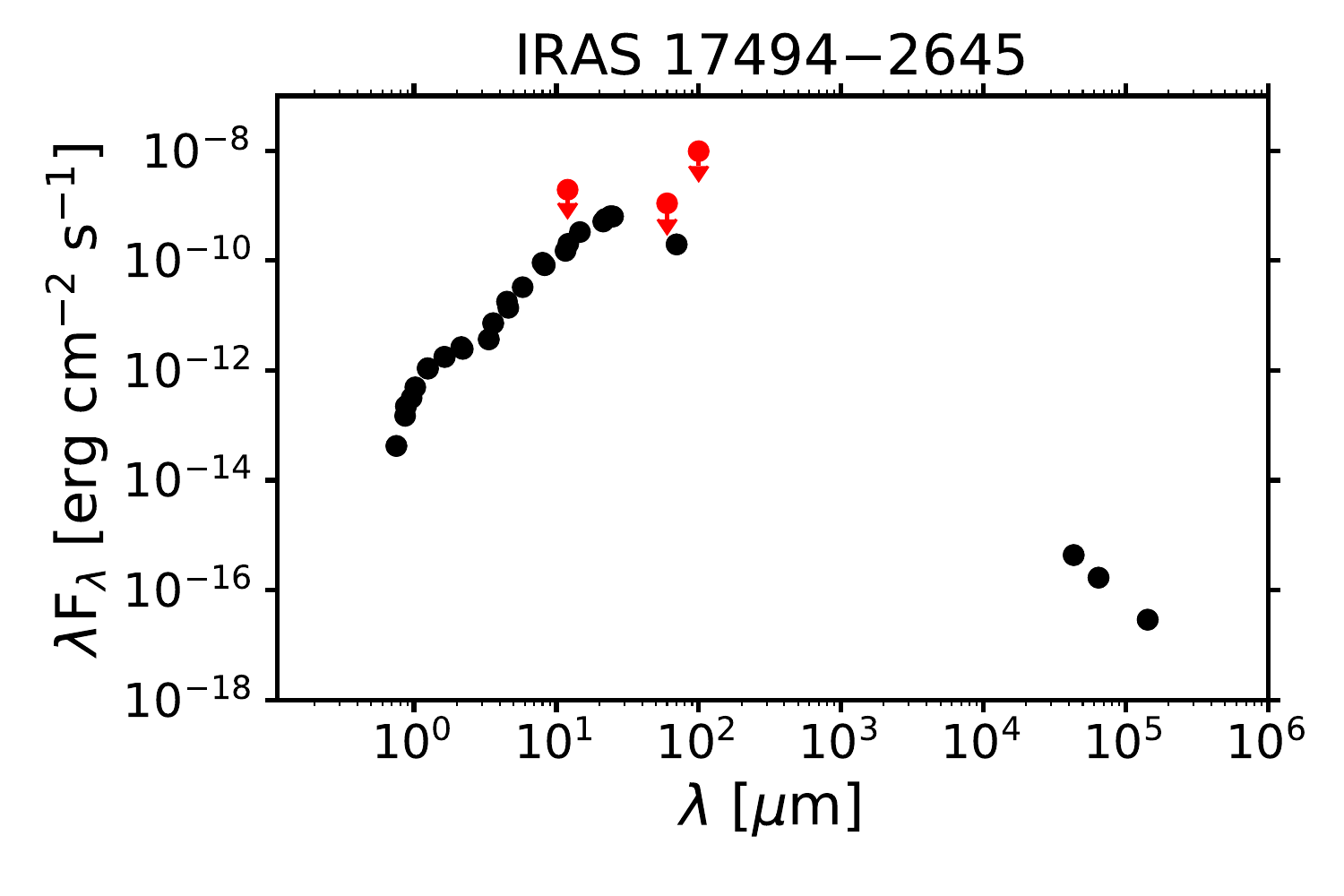}
	\includegraphics*[width=0.49\textwidth]{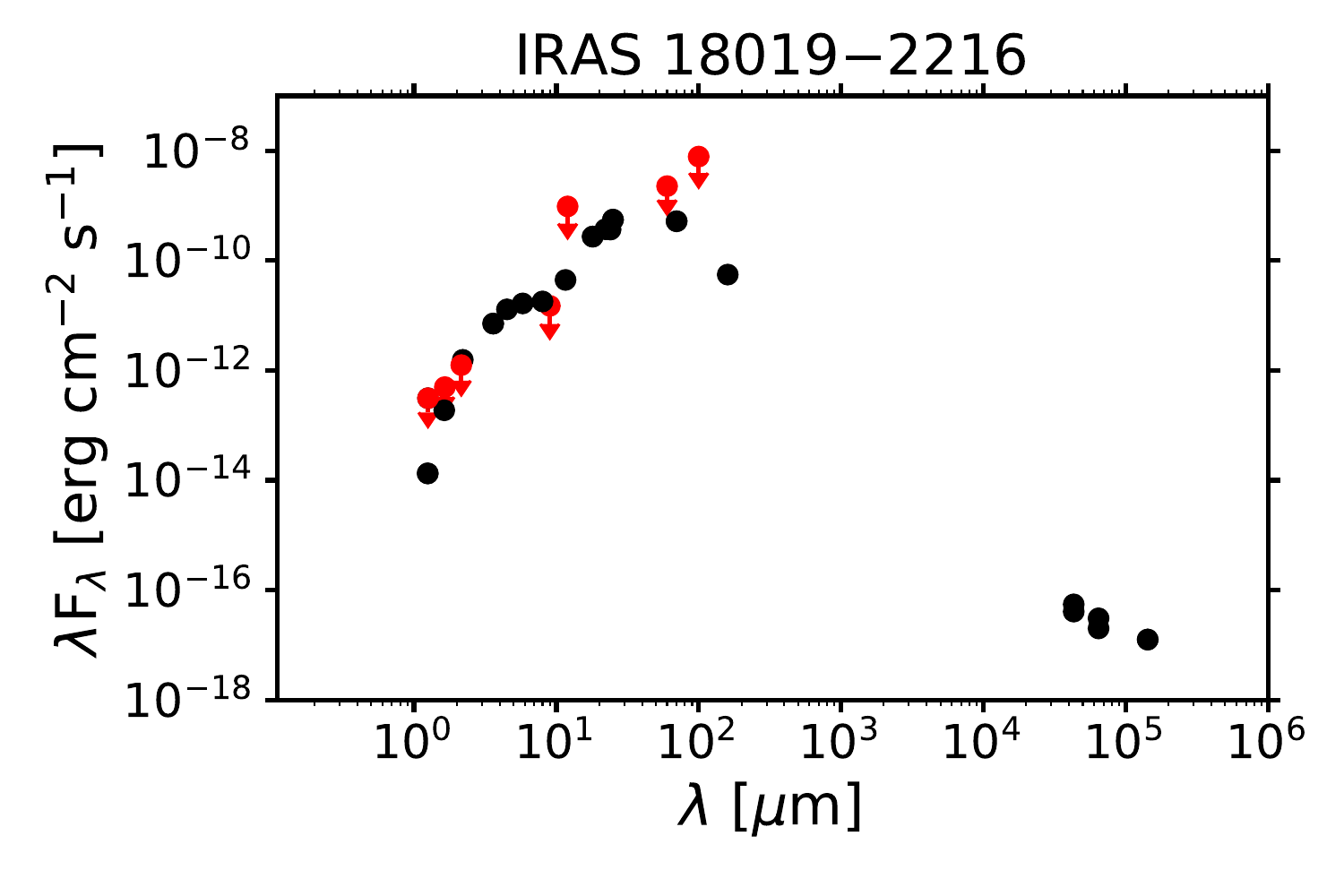}
	\caption{SED of the new OHPNe candidates, using optical and infrared photometric data mentioned in Section \ref{sec:red} and radio continuum emission from Table \ref{tab_fluxes}. Filled black circles are the  photometric measurements, while red symbols and arrows represent upper limits. }
	\label{fig_SED}
	
\end{figure*}

In order to better understand the circumstellar material surrounding these possible PNe, we obtained the optical-infrared SED of the four OHPNe candidates using archival data from $\sim$ 1 - 140 $\micron$. To this end, we gathered information from the Two Micron All Sky Survey (\textit{2MASS}), \textit{AKARI}, the Deep Near Infrared Survey of the Southern Sky (\textit{DENIS}), \textit{Gaia}, the \textit{Herschel Space Observatory}, Infrared Astronomical Satellite (\textit{IRAS}), the Midcourse Space Experiment (\textit{MSX}), the Panoramic Survey Telescope and Rapid Response System (\textit{Pan-STARRS}), the \textit{Spitzer Space Telescope}, the United Kingdom Infra-Red Telescope (\textit{UKIDSS}), the \textit{USNO-B} catalog, the Visible and Infrared Survey Telescope for Astronomy (VISTA) variable survey (\textit{VVV}) and the Wide-field Infrared Survey Explorer (all-\textit{WISE} survey). These photometric data are presented in Fig. \ref{fig_SED}, which includes also the radio continuum data previously plotted in Fig. \ref{fig:cont_spec}. 

The SEDs of all four objects are similar to each other, with a peak at $\lambda < 70$~$\micron$. This is also consistent with the SEDs observed in the rest of confirmed OHPNe \citep{usc12}, as well as in obscured  post--AGB stars and PNe, in general \citep[e.g.][]{ram09,ram12}. This peak of the SED is characteristic of PNe, rather than H\,{\sc ii} regions around massive YSOs, which also show thermal radio continuum emission but whose SED tends to peak at longer wavelengths \citep[$\ga 100$ $\micron$;][]{wc89,and12}. 

\section{Notes on the new OHPN candidates}
\label{sec:description_candi}

A summary of the radio continuum and OH maser parameters of the four OHPN candidates is given in Tables \ref{tab_fluxes} and \ref{tab_OH_sources}, respectively. In particular, Table \ref{tab_OH_sources} illustrate the number of OH maser components in each source, and the figure number in Qiao et al. papers where their OH spectra can be consulted. In this section, we provide a detailed description for each object.

\subsection{IRAS 16372--4808}\label{i16372}

The OH emission from this source was first reported  in the ATCA/VLA survey of \cite{sev97a}, showing a single-peaked spectrum with a flux density of $S_{\rm \nu}$ =  $2.25\pm 0.03$ Jy at 1612 MHz in 1994. Its flux density has remained relatively stable in subsequent observations \citep{deac04,qia16b}. No H$_{2}$O maser emission has been detected towards this source \citep{deac07}.  

The radio continuum emission in this object was first reported by \citet{urqu07}. As mentioned above, our results show that the radio continuum of this object has increased by a factor of $\sim 1.2$ in around 10 yr (Table \ref{tab_fluxes}).

\cite{bains09} noticed the spacial coincidence between radio continuum and OH maser emission using interferometric data. They argue that this object is a post--AGB star. This conclusion is based on an DUSTY modeling \citep{ive99} of its CSE, from which was obtained an $T_{\rm{eff}}\simeq$ 14,400 K for its central star, which is significantly lower than $\sim$25,000 K necessary to sustain a photoionized nebula. Thus, \cite{bains09} suggested that the radio continuum emission may arise from a  shock-ionized nebula. However, we note that DUSTY assumes spherical symmetry of the CSE, but post--AGB stars with radio continuum emission usually have non-spherical morphologies \citep[e.g. M 1--92, Hen 3--1475, the Red Rectangle, IRAS 15445--5449 or IRAS 18043--2116 in][respectively]{buj94, bor97, coh04, per11, per17}. Furthermore, the spatially resolved emission from radio, infared and/or optical observations of the OHPNe K 3-35, IRAS 17347--3139 and IRAS 16333--4807 show bipolar nebulae with a toroidal structure at their equatorial plane \citep[see][]{mir01, deG04, usc08, taf09, sansa12, usc14}. Therefore, the effective temperature obtained with DUSTY may not be accurate. Our results are compatible with the radio continuum emission arising from photoionized gas (see Sections \ref{sec:spec_ind} and \ref{sec:winds}).
 
\begin{figure*}
	\begin{centering}
		\includegraphics*[width=1\textwidth]{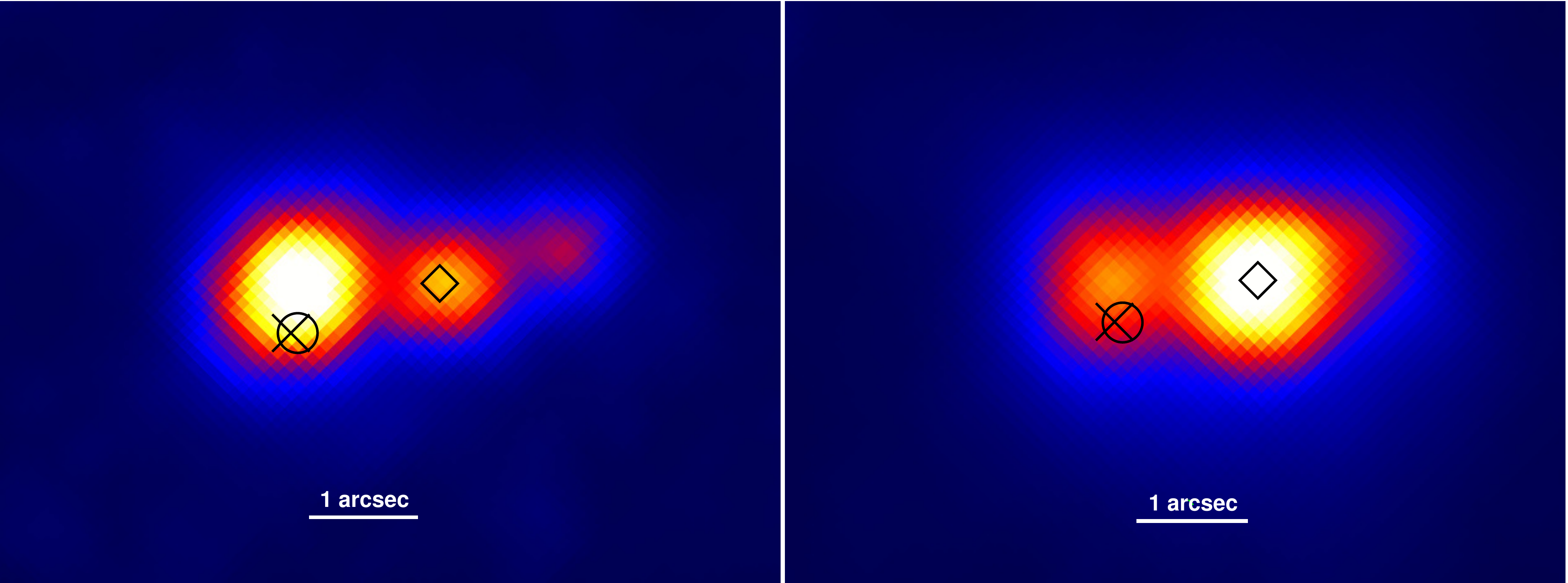}
		\caption{Infrared emission around IRAS 16372--4808 in \textit{Ks} (left) and \textit{J} (right) filters taken from the \textit{VVV} survey, both in logarithm flux colour scale. The circle and cross are centered on the positions of the radio continuum emission peak at 2.1 GHz and the 1612 MHz OH maser spot of IRAS 16372-4808, respectively. The diamond is centered on the GAIA's source (see text for details). North is up and East is to the left.} 
		\label{fig:I16372_infra}
	\end{centering}
\end{figure*}

We found a source (GAIA EDR3 5941005128417236224) in the GAIA EDR3 catalog \citep{gaia21}, whose reported position is  $\simeq 1.\arcsec4$ away from the radio continuum emission peak. Considering that this distance is small, but somehow larger than the estimated astrometric uncertainty of SPLASH--ATCA ($\simeq 1\arcsec $), the association of IRAS 16372--4808 with this optical source is uncertain. Furthermore, infrared \textit{J} and \textit{K} images from the \textit{VVV} survey (Fig. \ref{fig:I16372_infra}) show at least two distinct sources, separated by $\simeq 1.\arcsec4$. The source detected in GAIA (R.A. [J2000] = 16:40:55.690559 ($\pm 0.00008\arcsec$), DEC [J2000] --48:13:59.20725 ($\pm 0.00005\arcsec$) coincides with the western \textit{VVV} source, while the OH maser and radio continuum, as well as the mid and far infrared emission (e.g, all-WISE survey) is spatially coincident with the eastern source. Therefore, this OHPN candidate could be a different object from the one showing the optical counterpart. Alternatively, the two infrared sources may trace two lobes of the same PN, tracing a bipolar morphology. Spectroscopic observations will also be useful to ascertain the true optical/infrared counterpart of the OHPN candidate.

\subsection{OH 341.6811+00.2634}

The 1612 MHz OH maser emission was reported for the first time with the SPLASH--ATCA observations \citep{qia16b}. The spectrum of the OH maser emission is single-peaked, with $S_{\rm \nu}$ = 0.66 $\pm$ 0.07 Jy. No H$_2$O maser was detected in the H$_2$O Southern Galactic Plane Survey \citep[HOPS;][]{wal11,wal14}.

With the SPLASH--ATCA data we also report for the first time this object as a radio continuum-emitter, as well as the interferometric coincidence between the radio continuum source and the 1612 MHz OH maser emission (Fig. \ref{fig:corr}, top right).

No obvious optical counterpart can be identified in DSS images and in the GAIA EDR3 catalog. We found, however, an \textit{IRAS} source relatively close ($\sim$27$\arcsec$) to the position of the radio continuum peak at 2.1 GHz of OH 341.6811+00.2634. This source is IRAS 16483-4342. However, the flux densities reported for IRAS 16483-4342 (3.59 Jy and 191 Jy at 25 $\micron$ and 100 $\micron$, respectively) are higher than the ones reported in \textit{MIPSGAL} and by the \textit{Herschel} telescope (2.26 Jy and 2.04 Jy at 24 $\micron$ and 70$\micron$, respectively) at a location close to the radio continuum emission. Therefore, IRAS 16483-4342 and OH 341.6811+00.2634 seem to be two different sources.

This OHPN candidate is associated in SIMBAD with the infrared source SSTGLMC 341.6811+00.2634, and for which infrared images in \textit{K} band retrieved from the \textit{VVV} survey (spatial resolution $\sim$ 0.\arcsec42) show unresolved emission.

\subsection{IRAS 17494--2645}

Single-peaked 1612 MHz OH maser emission was discovered by \cite{sev97a}, showing $S_{\rm \nu}$ = 2.62 $\pm$ 0.03 Jy. The spectrum reported in SPLASH--ATCA by \cite{qia18} shows $S_{\rm \nu}$ = 3.05 $\pm$ 0.07 Jy. Variability in the maser emission cannot be confirmed because the spectral resolution of the spectrum presented in \cite{sev97a} is different from that of \cite{qia18} (1.36 km s$^{-1}$ and 0.09 km s$^{-1}$, respectively). No H$_2$O maser was detected in HOPS \citep{wal11,wal14}.

The radio continuum emission at 5 GHz on this source was detected in observations made in 1990 \citep{white05}. Its flux density seems to have increased by a factor of 2.0 $\pm$ 0.45 in around 26 yr, as mentioned above (Table \ref{tab_fluxes} and Section \ref{sec:spec_ind}). While both OH maser and radio continuum emission has been previously reported individually, our SPLASH--ATCA data allow us to accurately ascertain their mutual association (Fig \ref{fig:corr}, bottom left). 

No obvious optical counterpart can be identified in either DSS images or the GAIA EDR3 catalog. However, the radio continuum emission peak at 2.1 GHz of IRAS 17494--2645 coincides with a star-like object in PanSTARRS (ID: 75872681281354413), with a reported mean AB magnitude in filter \textit{i} (at $\sim$7545 \AA) of 21.33 $\pm$ 0.03 mag. Infrared emission from \textit{Ks} and \textit{J} filters taken from the \textit{VVV} survey (resolution of 0.\arcsec45 and 0.\arcsec70, respectively) show spatially unresolved emission from the object.

\subsection{IRAS 18019--2216}\label{i18019}

\cite{sev97a} reported the 1612 MHz OH maser spectrum with a single feature showing S$_{\rm \nu}$ = 1.33 $\pm$ 0.03 Jy at --31.8 $\pm$ 1.36 km s$^{-1}$. \cite{sev01} reported afterwards  S$_{\rm \nu}$ = 0.96 $\pm$ 0.03 Jy at --32.2 $\pm$ 1.36 km s$^{-1}$. With respect to these spectra, the spectrum presented in \cite{qia20} shows two new additional maser features. The brightest single feature shows S$_{\rm \nu}$ = 0.77 $\pm$ 0.07 Jy at --31.3 $\pm$ 0.1 km s$^{-1}$. With respect to this previously detected brightest feature, one of the two new observed maser spots is redshifted, whereas the other one is blueshifted. The former shows 0.74 $\pm$ 0.07 Jy at --35.6 $\pm$ 0.1 km s$^{-1}$, and the latter shows 0.23 $\pm$ 0.07 Jy at --27.2 $\pm$ 0.1 km s$^{-1}$. The three features show a rough north--south (N-S) distribution in Fig. \ref{fig:corr}, along a length of $\sim$ 1.$\arcsec$8. However, given the N-S elongation of the synthesized beams, this distribution is highly uncertain. High--angular resolution observations, with a better \textit{uv} coverage are needed to accurately determine the structure traced by the OH maser emission. No H$_2$O maser was detected in HOPS \citep{wal11,wal14}.

With the SPLASH--ATCA data we detected this object for the first time in the radio continuum, as well as we reported here for the first time interferometric association between the radio continuum and the OH maser (Fig \ref{fig:corr}, bottom right). We found IRAS 18019--2216 with an S/N $\simeq$ 2 at 2.1 GHz, and we were able to detected it in two different epochs from the VLA archive at 4.65 GHz and 6.95 GHz (see Table \ref{tab_fluxes}). 

We did not find an optical counterpart for this source. In SIMBAD it is catalogued as the YSO candidate SSTGLMC G007.9579--00.3931. Infrared images in \textit{K} band taken from the \textit{VVV} survey show spatially unresolved emission (resolution $\sim$ 0.\arcsec40). 

\begin{table}
	\caption{Properties of OH maser emission at 1612 MHz for the new OHPN candidates$^{a}$.}
	\label{tab_OH_sources}
	\begin{tabular}{lcccl}
		\hline
		\hline
		& & $\upsilon_{LSR}$$^{b}$ & $S_{\nu}$$^{c}$ \\
		Name		 & SPLASH--ATCA ID &	km s$^{-1}$ & Jy  & Figure $^{d}$   \\
		\hline
		IRAS 16372--4808 &  G337.064-1.173	& -90.5 & 1.94  & 1, 40 \\
		OH 341.6811+0.2634 & G341.681+0.264 & 54.0 & 2.16 & 1, 126\\
		IRAS 17494--2645 & G002.640-0.191 & 124.1 & 6.10 & 2, 295 \\
		IRAS 18019--2216 & G007.958-0.393 &  -31.3 & 2.09 & 3, 325 \\
			           & & -35.6 & 1.47 &  \\
			           & & -27.2 & 0.18 &  \\
		\hline
	\end{tabular}
	
	$^{a}$ Values listed correspond to the individual OH maser components identified in the SPLASH--ATCA observations by \citet{qia16b,qia18,qia20}\\
	$^{b}$ Local standard of rest velocity of the OH components. See the text for details.\\
	$^{c}$ Flux density of the OH components.\\
	$^{d}$ Figure number for the OH spectrum in the corresponding references: 1: \citet{qia16b}. 2: \citet{qia18}. 3: \citet{qia20}. \\

\end{table}

\section{Discussion}\label{sec:dis}

\subsection{The nature of the OHPN candidates as evolved stars}\label{sec_natS}

Our main criterion to look for new nascent PNe is based on determining a spatial coincidence between OH maser and radio continuum sources, using interferometric observations. However, there are other sources, different from PNe, which can show both types of emission. The first step to ascertaining the nature of these sources is to determine whether they are evolved stars, or if they could be YSOs. Massive YSOs can sustain photoionized regions around them. However, the properties of our OHPNe candidates are more consistent with their being stars in the last stages of stellar evolution. 

The first clue is the OH transition detected in these objects. While massive YSOs are commonly associated to the main-line OH masers at 1665 MHz and 1667 MHz transitions  \citep*[e.g.,][]{cas80,szy04},
our OHPNe candidates only show OH maser emission at 1612 MHz. This is a typical characteristic of OH masers in evolved stars. In fact, while stars from the AGB up to the PN phases can show also OH maser emission at other frequencies, the satellite line at 1612 MHz is generally brighter than the main-line OH masers in these type of objects \citep{lew97}. This is also the case of all confirmed OHPNe \citep{usc12,qia16a}. 

Line emission from different molecules can also provide further constraints. For instance, methanol maser emission is widespread in massive star-forming regions, often associated with interstellar OH masers \citep{men91}. Furthermore, methanol masers have never been detected in CSE \citep{breen13, gom14}. The position of all our sources was covered by the Methanol Multibeam Survey, but none was detected \citep{cas10,cas11}, again consistent with their being evolved stars. 

We have also inspected the public data from HOPS\footnote{\url{https://research.science.mq.edu.au/hops/public/data_cubes.php}}, which included cubes of the emission from inversion transitions of ammonia at $\simeq 23$ GHz \citep{pur12}. These transitions are tracers of gas with high density \citep[$>10^3$ cm$^{-3}$,][]{ho83}. Stars are born with dense cores in molecular clouds, and YSOs systematically host ammonia emission \citep[e.g.,][]{ang89,ver89}. The sources OH 341.6811+00.2634, IRAS 17494–2645, and IRAS 18019–2216 were within the sky section covered by HOPS, but no ammonia emission was found to be associated with them. The closest ammonia emission to these sources is associated with the compact dense core [PLW2012] G007.975-00.361+133.4, whose peak emission is located at $\sim$ 133 km s$^{-1}$ and $\simeq$2 arcmin away from IRAS 18019-2216, which rules out an association with any of the 1612 MHz OH maser spots and continuum emission in this object.

The SEDs of the OHPNe candidates (Section \ref{sec:red}, Fig. \ref{fig_SED}), with peaks at $\lambda < 70$~$\micron$ also suggest that they are evolved objects. On the other hand, in massive YSOs the peak emission is expected at $\lambda > 100$~$ \micron$ \citep[][]{wc89, and12}. We also note that the wavelength of this peak emission in their SEDs is also similar to the confirmed OHPNe \citep[][]{usc12}. Moreover, mid-infrared images from the \textit{Spitzer Space Telescope} show bright, spatially unresolved, isolated sources \citep[][]{qia16b,qia20}, while massive star--forming regions usually show extended infrared emission, with filaments and diffuse nebulosities. 

\begin{figure}
	\centering
	\includegraphics[width=0.5\textwidth]{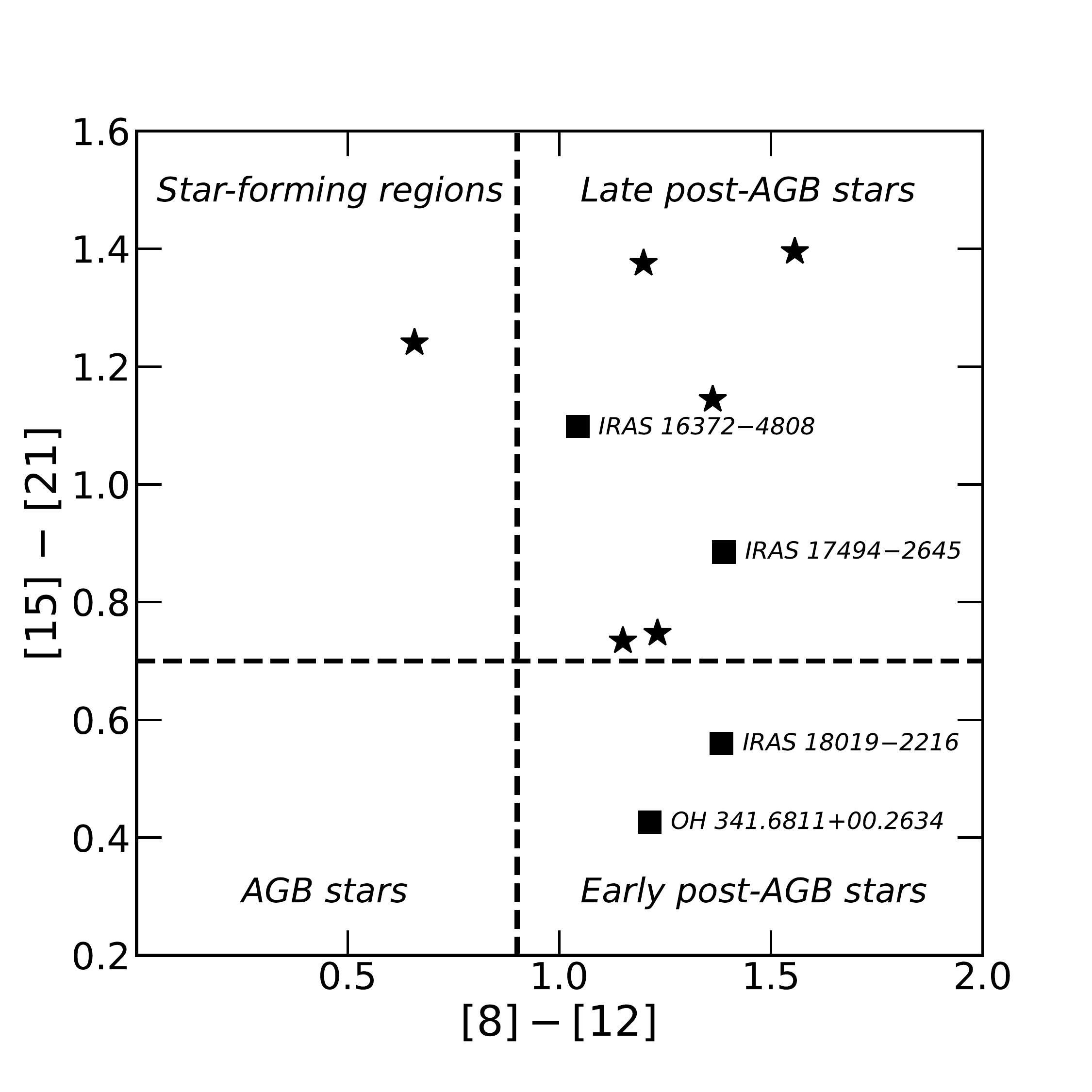}
	\caption{\textit{MSX} colour-colour diagram, as defined by \citet{sev02}. 
	The vertical and horizontal dashed lines separate the diagram into four quadrants, where different types of sources tend to cluster. The filled star and square symbols represent the known confirmed OHPNe and new candidates of SPLASH--ATCA, respectively.} 
	\label{fig:msx}
\end{figure}

A more detailed analysis of the infrared colours of the OHPNe candidates can shed further light on the nature of these sources. In Fig. \ref{fig:msx} we show an \textit{MSX} colour--colour diagram, defined as $[a]-[b]$ = 2.5 $log(S_{a}/S_{b})$, where $a$ and $b$ are wavelength in $\micron$ and $S$ is flux density in Jy, which is useful to discriminate the mid--infrared colours typical of star-forming regions and of evolved stars \citep{sev02}. The previously confirmed OHPNe and new candidates have similar colours, consistent with those in evolved stars, and in particular of those in the post--AGB phase. 

While none of the criteria above can provide by themselves an unquestionable determination of the nature of these sources as evolved stars, the combined evidence from them all makes it highly unlikely that these objects belong to a different category of sources.

\subsection{Radio continuum emission: photoionized nebulae or winds ionized by other processes?}
\label{sec:winds}

We have argued above that these sources have a high probability of being evolved stars. But the determination of whether these objects have reached the PN phase requires ascertaining whether photoionization of the CSE has already started. A solid classification as PNe would require optical/infrared spectroscopy, which is not available for our candidates. However, the radio continuum emission can also provide important clues on the nature of these objects. In particular, we can study whether this emission is free-free radiation from ionized gas and, if so, whether the ionization is produced by UV radiation from the central star (i.e., it is in the PN phase), or by other processes, such as shocks in high-velocity winds (which can also be produced previously, in the AGB or post--AGB phases), with no significant contribution from photoionization. It is important to note that PNe can show both slowly expanding gas and high-velocity jets. Our discussion in this section, however, would focus on whether or not the mass--loss rates derived under the assumption of winds (ionized by shocks or other processes different from photoionization from the central star) are within reasonable values for objects in evolutionary stages before the PN phase.

As mentioned in Section \ref{sec:spec_ind}, the spectral indices of the radio continuum emission in these OHPNe candidates are consistent with free-free radiation from photoionized gas. Even in some cases, the spectral indices are close to 0.6, which suggests a radial dependency of electron density as $n_e\propto r^{-2}$, as one would expect in the photoionized region in a nascent PNe. However, we note that the same spectral dependency would be obtained in the case of free-free radiation from an ionized wind \citep{pan75,reyn86}, which may be produced in the post--AGB phase, before the photoionization of the CSE takes place. This situation is seen in jets from YSOs \citep{ang18}. Low-mass YSOs do not produce enough ionizing radiation to maintain an H\,{\sc ii} region around them, but radio continuum emission with spectral indices $\simeq 0.6$ are detected from collimated jets. In that case, the emission may arise from shock-ionized material.

Shock-ionized outflows can thus produce exactly the same radio spectral indices as a photoionized region, regardless of whether they represent an isotropic or a collimated mass--loss. It is mainly the radial dependency of the electron density which determines the spectral index. However, the values of flux density can be used to constrain the nature of the emission.

For instance, the radio continuum emission from an isotropic ionized outflow would have $\alpha = 0.6$, and an $S_{\rm{\nu}}$ at a distance $d$ from the Sun \citep{pan75}: 

\begin{equation}
	\begin{aligned}
		S_{\rm{\nu}} = 5.12 \bigg[ \frac{\nu}{10 \hspace{0.07cm} \rm{GHz}} \bigg]^{0.6} \bigg[ \frac{T_{e}}{10^{4} \hspace{0.07cm} \rm{K}}\bigg]^{0.1} \bigg[ \frac{\dot{M}}{10^{-5}\hspace{0.07cm} \rm{M_{\odot}\hspace{0.07cm}yr^{-1}}} \bigg]^{4/3} \bigg[\frac{\mu}{1.2} \bigg]^{-4/3} \\
		\bigg[\frac{\upsilon_{\rm{exp}}}{\rm{10^{3} \hspace{0.07cm}km \hspace{0.07cm}s^{-1}}}\bigg]^{-4/3} \bar{Z}^{-2/3} \bigg[\frac{d}{\rm{kpc}}\bigg]^{-2}
	\end{aligned}
\end{equation}

The determination of mass--loss rates (under the assumption that the observed radio continuum emission arises from a shock-ionized wind) is hampered by the absence of accurate distance estimates for these sources, although we can still check whether the observed flux densities give reasonable values of those mass--loss rates.
Using the flux densities at the measured frequency closest  to 10 GHz for each source (see Table \ref{tab_fluxes}), an electron temperature of $T_{\rm e} = 10^{4}$ K, $\mu = 1.2$ for the mean atomic weight per electron, $\upsilon_{\rm{exp}}$ = 1 $\times$ 10$^{3}$ km s$^{-1}$ for the velocity of a post--AGB fast wind, and $\bar{Z}$ = 1 for metallicity, hence the mass--loss rate expected for the new OHPNe candidates would be $\dot{M}$ = 7.4 $\times$ 10$^{-5}$ (IRAS 16372--4808), 1.3 $\times$ 10$^{-6}$ (OH 341.6811+00.2634), 8.5 $\times$ 10$^{-6}$ (IRAS 17494--2645), 1.2 $\times$ 10$^{-7}$ (IRAS 18019--2216) $\rm{M_{\odot}}$ yr$^{-1}$ $(d/\rm{kpc})^{3/2}$.

On the other hand, as mentioned above, a collimated, biconical ionized jet would also give a spectral index ($\alpha = 0.6$). In this case, departures from this spectral index can be interpreted as deviations from the biconical symmetry \citep{reyn86}. There is ample evidence for the presence of collimated winds in post--AGB stars, and they can play a key role in PN shaping \citep{sah98}. For a collimated jet, the mass--loss rates required to explain the observed radio continuum emission, would be around one order of magnitude smaller than for the spherical ionized wind case, depending on the inclination of the jet with respect to the observer and, specially, on the degree of collimation of the jet \citep[][]{reyn86}.

In the case of our candidate OHPNe, the mass--loss rates required to explain the observed radio continuum emission as shock-excited winds are probably too high when comparing with the values expected in classical post--AGB models  \citep[$10^{-7}$--10$^{-8}$ $\rm{M_{\odot}}$ yr$^{-1}$;][]{vass94,block95,mil16}, but not so much if we consider the case of binary/multiple stellar systems. In these systems, large accretion rates between companions could be produced, with processes such as a common envelope \citep{iva13}. These processes can lead to extremely high mass--loss rates, as in the case of WF AGB and post--AGB stars \citep{imai07}, with total mass--loss rates of $\simeq 10^{-3}-10^{-4}$ M$_\odot$ yr$^{-1}$ \citep{riz13, kho21}. Thus, mass--loss rates from post--AGB stars in binary/multiple systems can be several orders of magnitude higher than rates expected from single stars. For instance, \citet{boll21} estimate mass--loss rates of $10^{-4}-10^{-8}$ $\rm{M_{\odot}}$ yr$^{-1}$ in a sample of jets in post--AGB binary systems.

We note, however, these large mass--loss rates at the AGB and post--AGB phases include mostly neutral material, which would not emit free-free radiation at radio wavelengths. The fraction of ionized material in winds and outflows from post--AGB stars is relatively low. For instance, while WFs are known to eject powerful collimated jets, their radio continuum emission is either extremely weak \citep[cf. non-detections at 22 GHz with  upper limits $30-40$ $\mu$Jy in several of these objects,][]{gom17}, or of non-thermal nature \citep{per13,sua15}. The WF IRAS 18043$-$2116 does have detectable radio continuum emission. From our SPLASH--ATCA data, we obtain a flux density of $\simeq 0.9$ mJy at 2.1 GHz, although the image is heavily contaminated by sidelobes from other sources in the field. A more reliable value of $\simeq 0.6$ mJy at 2 GHz was obtained by \citet{per17}. These authors also showed that the spectral index of the radio continuum emission is consistent with free-free emission from an ionized jet. Their derived mass--loss rate under the assumption that the radio continuum emission arises from an ionized wind is $\la \times 10^{-5}$ $\rm{M_{\odot}}$ yr$^{-1}$, more than two orders of magnitude lower than the mass--loss rate derived by \citet{kho21} from molecular line observations. In the case of the post--AGB binary system TW Cam, a source for which \citet{boll21} derived a mass--loss rate in its jet ($2\times 10^{-4}$ M$_\odot$ yr$^{-1}$), it is undetected in the NRAO VLA Sky Survey \citep[NVSS;][]{con98}, with a 3$\sigma$ upper limit of $\simeq 1.5$ mJy at 1.4 GHz, and in the VLA Sky Survey \citep[VLASS;][]{vlass21}, with a 3$\sigma$ upper limit of $\simeq 0.4$ mJy at 3 GHz, while the source is at a distance of 1.8 kpc, probably much closer than our OHPN candidates. There is, of course, a caveat in this flux density comparison, since the distance to our OHPN candidates is unknown, but these objects are not conspicuously bright in the infrared (Fig. \ref{fig_SED}) compared with other post--AGB stars \citep[see e.g., the SEDs in][]{ram12}, and therefore, they do not seem located particularly nearby. Moreover, it is likely that at least some of these objects are in the Galactic Bulge, given their coordinates, which would place them at $\simeq 8\pm 2$ kpc.

Other post--AGB stars with radio continuum emission also show weaker emission than these OHPN candidates. For instance, Hen 3-1475 displays a flat spectrum, with flux densities 0.3 -- 0.4 mJy between 2.6 and 8.6 GHz \citep{cer17}, while the Red Rectangle has a flux density of $\simeq 0.5$ mJy at 8.4 GHz \citep{kna95,jur97}. The object M 1--92 has radio fluxes comparable to those of our candidates \citep[$2.1\pm 0.4$ mJy at 3 GHz;][]{vlass21}, although it is relatively nearby \citep[2.5 kpc;][]{buj97}, consistent with its higher infrared emission (e.g., IRAS flux of 17.5 Jy at 12 $\mu$m, compared with 2 Jy for IRAS 16372-4808, our brightest object). 

In summary, considering that the radio continuum emission of the new OHPNe candidates tends to be stronger than in post--AGB stars (e.g. Hen 3--1475, M 1--92 or the Red Rectangle), including those showing strong shocks (e.g. the WFs), we favour that this continuum emission is likely to arise from photoionized gas, especially for the sources with brighter radio flux densities. Therefore, the radio continuum detected in the OHPN candidates is indicative of fossil shells ionized by UV photons from the central star of a PN, and other sources of excitation seem to be improbable.

\subsection{Comparison with known OHPNe and a possible evolutionary sequence}

As shown above, the radio continuum emission alone cannot provide unambiguous evidence that these objects have already entered the PN phase. However, their common characteristics with the confirmed OHPNe can provide further support of their likely nature as PNe.

Three of our candidates show OH maser lines at 1612 MHz with a single spectral feature. Only IRAS 18019--2216 shows multiple spectral components with flux densities of the same order, although previous observations also reported a single peak.
A dominant spectral feature can also be considered a common characteristic of known OHPNe. The OH spectra in these sources \citep{gom16,qia16a} are irregular, but they tend to be dominated by several blended features that would appear as a single peak when observed with limited spectral resolution. Additional, spectrally resolved components are much weaker than the main features. The case of multiple, well-separated OH maser components in IRAS 18019--2216 could be considered similar to that of K 3-35, although this characteristic seems to be less frequent among OHPNe. In any case, the characteristics of OH maser spectra in our candidate sources are different from the double horned profile observed on AGB stars such as OH/IR stars that typically traces the receding front and back sides \citep{reid76}, as well as the terminal expansion velocity of the CSE \citep[$\simeq 15-30$ km s$^{-1}$;][]{telin91}.

Another potentially interesting characteristic is the variation of the radio continuum flux density. Among the three cases in which we report data in more than one epoch (Table \ref{tab_fluxes}), there seems to be an increase of the emission in IRAS 16372--4808 and IRAS 17194--2645. In IRAS 18019--2216, the differences in flux density are not statistically significant. An enhancement of flux density has also been reported in the OHPNe Vy 2--2 and IRAS 17347--3139 \citep{chr98,gom05,taf09}, and interpreted as a sign of the growth of the photoionized region, as expected in a nascent PN while the ionization front proceeds along the CSE. In addition to maser-emitting PNe, a similar increasing trend has been observed in other young PNe \citep[e.g.,][]{kwo81a, kna95}. 
In contrast, non-thermal radio continuum emission in evolved stars shows variability that could be either in a increasing or decreasing direction \citep{per13,sua15,cer17}. In this sense, finding a decreasing flux density with time would be indicative of the presence of a non-thermal emission process. In the case of our candidates, the  enhancement of radio continuum emission is consistent with a nature as nascent PNe.

As discussed in Sections \ref{sec:spec_ind} and \ref{sec:winds} the spectral indices in our four OHPNe candidates are compatible with the ones expected in nascent PNe. This is further illustrated in Table \ref{tab_indices}, where we show the radio spectral indices of the previously reported OHPNe together with those of the candidates presented here. As mentioned before, a photoionized region would show free-free emission with spectral indices between 2 (optically thick) and -0.1 (optically thin). The optical depth of free-free emission decreases with frequency, so the same object can display both regimes, separated at a turnover frequency. We also mentioned that objects with a radial dependency of electron density of $n_e\propto r^{-2}$ can show a partially optically thick spectrum with $\alpha\simeq 0.6$ over a wide range of frequencies. It suggests a possible evolutionary sequence of nascent PNe based on their spectral indices. In this sense, the beginning of photoionization could be characterized by steep spectral indices (close to 2), when the densest parts of the CSE are photoionized. As the ionization front evolves, and the emission arises from regions both optically thin and thick, the resulting spectral indices would tend to be $\alpha \simeq 0.6$. The free-free emission would then become optically thin at high frequencies (with $\alpha\simeq -0.1$), with a turnover frequency that would decrease with time \citep{kwo81b}. In general, we would expect that the radio continuum spectrum would flatten with time over a wider range of frequencies. 

While this could be a general trend, it is difficult to set up a precise ranking of ages among confirmed and candidate OHPNe in Table \ref{tab_indices}, considering other possible parameters (e.g. the $M_{\rm i}$ of the stars, or binary evolution), and that the frequency coverage of the radio data is not homogeneous. However, we would suggest that sources like Vy 2--2 and IRAS 17494--2645 could be the youngest among the PNe in Table \ref{tab_indices}, since their spectral indices are steeper over a large frequency range, while K 3--35 and JaSt 23 would be relatively more evolved, since their radio spectra become optically thin within the range of sampled frequencies. A detailed and homogeneous study of the radio spectrum of these sources, as well as a monitoring of the radio continuum emission would be useful to better understand their evolutionary status. 

\begin{table}
	\caption{Radio spectral indices of the confirmed OHPNe and detected candidates.}
	\label{tab_indices}
	\begin{tabular}{lrrrr}
		\hline
		\hline
		& $S_{\nu} \propto \nu^{\alpha}$ & Frequencies & Confirmed \\
		Source name			 & 	$\alpha$		& GHz &  &   \\
		\hline
		JaSt 23 & 0.57 $\pm$ 0.24 $^{a}$ & 2.1 -- 5 & Yes &  \\
		& 0.09 $\pm$ 0.25 & 5 -- 10  &   \\
		K 3--35 & 0.6 $\pm$ 0.1$^{b}$ & 1.5 -- 10  & Yes\\
		& -0.1 $\pm$ 0.2 & 10 -- 22  &   \\
		IRAS 17393--2727 & 0.66 $\pm$ 0.09$^{c}$ & 2.1 -- 15 & Yes  \\
		IRAS 16333--4807 & 0.69 $\pm$ 0.01$^{d}$ & 2.1 --22.2 & Yes \\
				& 1.39 $\pm$ 0.08$^{e}$ & 1.38 -- 2.41 &   \\
		OH 341.6811+00.2634 & 0.77 $\pm$ 0.07 & 2.1 -- 45 &  No \\
		IRAS 17347--3139 & 0.79 $\pm$ 0.04$^{f}$ & 4.3 -- 8.9 & Yes  \\
		& 0.64 $\pm$ 0.06 & 16.1 -- 24 &   \\
		IRAS 16372--4808 & 0.90 $\pm$ 0.03 & 2.1 -- 9.7 & No   \\
		& 0.375 $\pm$ 0.008 & 8.7 -- 45 &   \\
		IRAS 18019--2216 & 1.0 $\pm$ 0.3 & 2.1 -- 6.5 & No \\
		Vy 2--2 & 1.30 $\pm$ 0.07$^{g}$ & 1.5 -- 15 & Yes   \\	
		IRAS 17494--2645 & 1.35 $\pm$ 0.09 & 2.1 -- 6.5 & No \\
		\hline
	\end{tabular}
	
	$^{a}$ Using our data $S_{\rm \nu}$ = 2.25 $\pm$ 0.4 mJy at 2.1 GHz, and $S_{\rm \nu}$ = 3.7 $\pm$ 0.4 mJy and $S_{\rm \nu}$ = 3.9 $\pm$ 0.4 mJy at 5 and 10 GHz from \cite{van01}, respectively.\\
	$^{b}$ \cite{aa93}.\\ %1988-Dec
	$^{c}$ Using at 2.1 GHz the arithmetic mean (weighted by the inverse of noise squared) of our data and that of \citet{gom16}; at 3 GHz the value is obtained from \citet{vlass21}; at 5 GHz and 15 GHz from \citet{pott87}.\\
	$^{d}$ Using our data $S_{\rm \nu}$ = 19.10 $\pm$ 0.54 mJy at 2.1 GHz, and $S_{\rm \nu}$ = 97 $\pm$ 10 mJy at $\sim$22.24 GHz from \cite{usc14}. \\
	$^{e}$ \cite{qia16a}. \\
	$^{f}$ \cite{gom05}. For the central region of this PN, \cite{taf09} obtained spectral indices of $\sim$1.34 and $\sim$0.90 between 8.4 -- 22.2 GHz, and 22.2 -- 43 GHz, respectively.\\
	$^{g}$ Using the most recent data in \cite{chr98}. Spectral indices of 1.43 $\pm$ 0.07, 1.28 $\pm$ 0.06, 1.24 $\pm$ 0.05 and 1.30 $\pm$ 0.07 are obtained from observations performed in 1982, 1987, 1992 and 1997, respectively.\\

\end{table}

\subsection{New colour diagnostics for OHPNe}

\begin{figure*}
	\centering
	\includegraphics[width=1\textwidth]{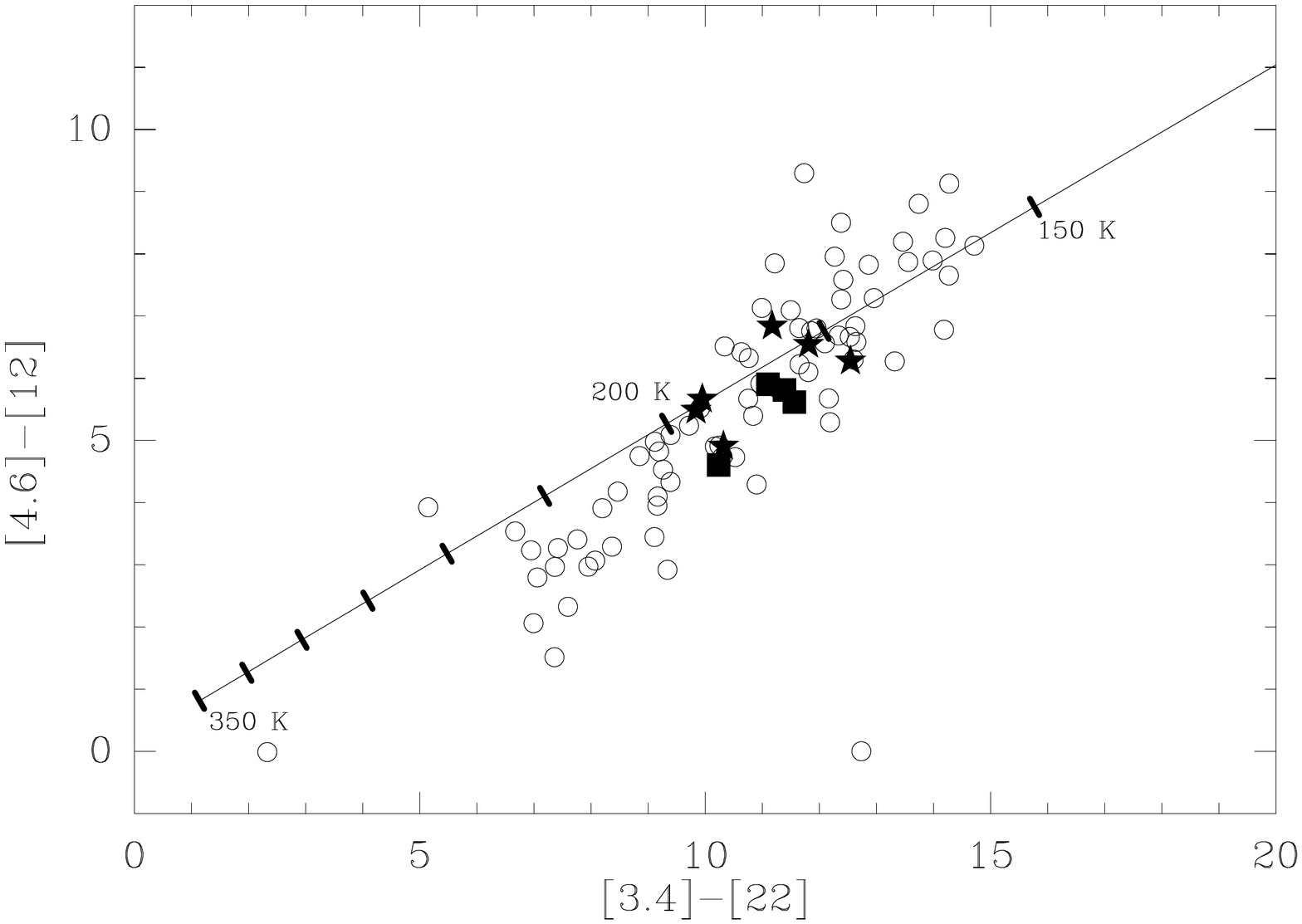}
	\caption{\textit{WISE} colour-colour diagram \citep[similar to the one presented in][]{gom17}. The stars and squares represents the confirmed OHPNe and the new candidates of SPLASH--ATCA, respectively. The open circles represent obscured post--AGB candidates \citep{ram09,ram12}. The solid line represents the locii of the colours for blackbody brightness distributions. The tick marks in the blackbody line go from 150 to 350 K at increment steps of 25 K. For IRAS 18019--2216, we have used measurements at 3.6 and 4.5 $\micron$ reported by \textit{Spitzer} because at these wavelengths, the images of \textit{WISE} show that the emission of this OHPN candidate is contaminated by a nearby source.}
	\label{fig:wise}
\end{figure*}

In Fig. \ref{fig:wise} we show a \textit{WISE} colour-colour diagram, where each colour is defined as $[a]-[b]$ = 2.5 $log(S_{a}/S_{b})$, and where we compare the infrared colours of confirmed OHPNe and our new OHPNe candidates, with optically obscured post--AGB candidates from \citet{ram09,ram12}. A similar diagram was presented in \cite{gom17}, but comparing the obscured post--AGB candidates with WF objects. These authors did not find any significant difference in colours between WFs and obscured post--AGB stars.

However, in the case of our colour-colour diagram, the OHPNe (both candidate and bona-fide), seem to cluster together around $9.5\la [3.4]-[22]\la 13.5$, and $4.0\la [4.6]-[12] \la 7.0$. 

The yet scarce number of known and candidate OHPNe precludes us to obtain strong statistical constraints, but a Kolmogorov-Smirnov test on the WISE colours suggests that the colour distribution between OHPNe and obscured post--AGB stars could be different ($p$ value $\simeq 0.16$, is the probability to obtain the observed $[3.4]-[22]$ colours assuming that OHPNe and post--AGB stars are drawn from populations with the same colour distribution). If these different colours are confirmed, this would suggest that OHPNe are a particular type of objects, and not all evolved objects go through this phase, a particularity that is not observed in WF objects. This may also suggest that WFs are not the direct precursors of OHPNe, despite they share the property of being maser-emitting evolved objects, and that all, except three WFs, are also OH emitters \citep{gom17}. 

In any case, this apparent clustering of OHPNe in the WISE colour-colour diagram can be used as a diagnostic to support that our candidates are indeed bona fide OHPNe, since they have similar colours as the confirmed objects in this class. Moreover, this \textit{WISE} colour-colour diagram could be used to identify more OHPNe candidates in the future.

\section*{Conclusions}

We have used data from the ATCA follow-up of the SPLASH survey to identify new candidates to being PNe with OH maser emission. OHPNe seem to be extremely rare, consistent with their proposed nature as nascent PNe. So far, only 6 of such OHPNe have been confirmed in the literature. Our criterion to identify new OHPNe was to find a spatial coincidence between radio continuum at 2.1 GHz and OH maser emission, in objects classified in SPLASH as evolved stars and sources with unknown nature. 

With these data, we have found four new candidates to being OHPNe: three in which we report the association between continuum and OH maser emission for the first time (OH 341.6811+00.2634, IRAS 17494--2645 and IRAS 18019--2216), and another one (IRAS 16372--4808) whose nature as a PN had been previously discarded. However, we discuss that there is not enough evidence for such a rejection, and should also be considered a OHPN candidate.

The radio continuum emission in all four candidates has spectral indices compatible with partially optically thick free-free emission from ionized gas and, in some cases, the spectral index is close to 0.6, the value expected from a ionized region with a radial dependency of density as $r^{-2}$, as it could be the case in nascent PNe.

Although massive YSOs can sustain photoionized regions (thus emitting free-free radio continuum emission) and present OH maser emission, these OHPN candidates are most likely evolved stars, based on the detection of OH ground-state maser emission only at 1612 MHz, the absence of methanol masers and of high-density molecular line tracers, their SED peaking at $\lambda < 70$ $\mu$m, and their infrared colours.

We also discuss whether the radio continuum emission actually arises from the photoionized region in a nascent PN, or from a shock-ionized wind (and therefore, the sources may still be in the post--AGB phase). While the mass--loss rates assuming an ionized wind are too high comparing with the expected ones in single post--AGB stars, we cannot completely discard that these rates could be enhanced by processes in binary/multiple systems, such as common envelope evolution. We suggest, however, that this is unlikely, since other post--AGB sources that have extremely high mass--loss rates show none or very weak radio continuum emission. Therefore, the radio continuum detected in the OHPN candidates is indicative of fossil shells ionized by UV photons from the central star of a PN. Their final confirmation as bona fide PNe would require optical/infrared spectroscopy.

However, these OHPN candidates show other similarities with confirmed OHPNe, such as the shape of the OH maser spectrum (in most cases dominated by a  single spectral feature), or the increase of radio continuum flux density with time. Spectral indices of radio continuum emission are also similar to confirmed OHPNe. We propose a general trend, with younger OHPNe having steeper spectral indices, that would flatten as the object evolves. Moreover, we find evidence that OHPNe (including our candidates) tend to cluster in a definite area of a colour-colour diagram using WISE data, with $9.5\la [3.4]-[22]\la 13.5$, and $4.0\la [4.6]-[12] \la 7.0$, compared with the wider spread of obscured post--AGB stars in the same diagram. We also suggest that this colour diagram could be useful to identify new OHPNe in the future.

\section*{Acknowledgements}
We thank our anonymous referee for his/her comments that have significantly improved the original manuscript. The Australia Telescope Compact Array is part of the Australia Telescope National Facility (grid.421683.a) which is funded by the Australian Government for operation as a National Facility managed by CSIRO.  We acknowledge the Gomeroi people as the traditional owners of the Observatory site. The National Radio Astronomy Observatory is a facility of the National Science Foundation operated under cooperative agreement by Associated Universities, Inc. 
We used continuum images from the VLA Sky Survey, downloaded from the Canadian Initiative for Radio Astronomy Data Analysis (CIRADA), which is funded by a grant from the Canada Foundation for Innovation 2017 Innovation Fund (Project 35999), as well as by the Provinces of Ontario, British Columbia, Alberta, Manitoba and Quebec.
This work has made use of the SIMBAD database, operated at the CDS, Strasbourg, France, and the NASA/IPAC Infrared Science Archive, which is operated by the Jet Propulsion Laboratory, California Institute of Technology, under contract with the National Aeronautics and Space Administration. It also makes use of data products from 2MASS (a joint project of the University of Massachusetts and the Infrared Processing and Analysis Center/California Institute of Technology, funded by NASA and the NSF), AKARI (a JAXA project with the participation of ESA), 
 DENIS (partly funded by the SCIENCE and the HCM plans of the European Commission under grants CT920791 and CT940627), the services of the ESO Science Archive Facility, \textit{Gaia} (the European Space Agency (ESA) mission {\it Gaia} (\url{https://www.cosmos.esa.int/gaia}), processed by the {\it Gaia} Data Processing and Analysis Consortium (DPAC, \url{https://www.cosmos.esa.int/web/gaia/dpac/consortium}). Funding for the DPAC has been provided by national institutions, in particular the institutions participating in the {\it Gaia} Multilateral Agreement), HERSCHEL (Herschel is an ESA space observatory with science instruments provided by European-led Principal Investigator consortia and with important participation from NASA), IRAS (was a joint project of the US, UK and the Netherlands), Pan-STARRS (have been made possible through contributions by the Institute for Astronomy, the University of Hawaii, the Pan-STARRS Project Office, the Max-Planck Society and its participating institutes), \textit{Spitzer} Space Telescope (operated by the Jet Propulsion Laboratory, California Institute of Technology under a contract with NASA),  MSX (funded by the Ballistic Missile Defense Organization with additional support from NASA Office of Space Science), UKIDSS (The project is defined in \cite{law07} and uses the UKIRT Wide Field Camera \citep[WFCAM;][]{cas07} and a photometric system described in \cite{hew06}. The pipeline processing and science archive are described in \cite{ham08}), USNO-B catalog (the construction and contents of the catalog can be found in \cite{mon03}), VVV survey (is supported by the European Southern Observatory, by BASAL Center for Astrophysics and Associated Technologies PFB-06, by FONDAP Center for Astrophysics 15010003, by the Chilean Ministry for the Economy, Development, and Tourism’s Programa Iniciativa Científica Milenio through grant P07-021-F, awarded to The Milky Way Millennium Nucleus), and WISE (a joint project of the University of California, Los Angeles, and the Jet Propulsion Laboratory/California Institute of Technology, funded by the NASA).
RC, JFG, and LFM acknowledge support from grant PID2020-114461GB-I00 and the ``Center of Excellence Severo Ochoa'' award for the Instituto de Astrof\'{\i}sica de Andaluc\'{\i}a (SEV-2017-0709), funded by MCIN/ AEI /10.13039/501100011033. RC is also supported by the predoctoral grant PRE2018-085518, funded by MCIN/AEI/ 10.13039/501100011033 and by ESF Investing in your future. This work is also partially supported by grants R18-RT-3082 (IAA4SKA) and P20-00880, funded by the Economic Transformation, Industry, Knowledge and Universities Council of the Regional Government of Andalusia and the European Regional Development Fund from the European Union.
LU acknowledges support from the University of Guanajuato (Mexico) grant ID CIIC 164/2022. HI has been supported by JSPS KAKENHI (25610043, JP16H02167). H.-H.Q. is partially supported by the Special Funding for Advanced Users, budgeted and administrated by Center for Astronomical Mega-Science, Chinese Academy of Sciences (CAMS-CAS), CAS ``Light of West China'' Program and the National Natural Science Foundation of China (grant No. 11903038).

%%%%%%%%%%%%%%%%%%%%%%%%%%%%%%%%%%%%%%%%%%%%%%%%%%
\section*{Data Availability}

%The inclusion of a Data Availability Statement is a requirement for articles published in MNRAS. Data Availability Statements provide a standardised format for readers to understand the availability of data underlying the research results described in the article. The statement may refer to original data generated in the course of the study or to third-party data analysed in the article. The statement should describe and provide means of access, where possible, by linking to the data or providing the required accession numbers for the relevant databases or DOIs.

All data at radio wavelengths presented in this paper are publicly accessible through the Australia Telescope Online Archive (\url{http://atoa.atnf.csiro.au}) and the NRAO Data Archive (\url{https://archive.nrao.edu/}), as uncalibrated visibilities. Relevant project numbers are listed in Table \ref{tab_archive}. The optical and infrared data we used are also accessible through their respective archives.



\bibliographystyle{mnras}
%\bibliography{OHPNe} % if your bibtex file is called example.bib

\begin{thebibliography}{99}
\bibitem[\protect\citeauthoryear{Aaquist}{1993}]{aa93}
Aaquist O.~B., 1993, A\&A, 267, 260
\bibitem[\protect\citeauthoryear{Aaquist \& Kwok}{1991}]{aa91} 
Aaquist O.~B., Kwok S., 1991, ApJ, 378, 599
\bibitem[\protect\citeauthoryear{Anderson et al.}{2012}]{and12}
Anderson L.~D., Zavagno A., Barlow M.~J., Garc{\'i}a-Lario P., Noriega-Crespo A., 2012, A\&A, 537, A1
\bibitem[\protect\citeauthoryear{Anglada et al.}{1989}]{ang89} 
Anglada G., Rodriguez L.~F., Torrelles J.~M., Estalella R., Ho P.~T.~P., Canto J., Lopez R., et al., 1989, ApJ, 341, 208
\bibitem[\protect\citeauthoryear{Anglada, Rodr{\'\i}guez, \& Carrasco-Gonz{\'a}lez}{2018}]{ang18} Anglada G., Rodr{\'\i}guez L.~F., Carrasco-Gonz{\'a}lez C., 2018, A\&ARv, 26, 3
%doi:10.1007/s00159-018-0107-z
\bibitem[\protect\citeauthoryear{Bains et al.}{2009}]{bains09}
Bains I., Cohen M., Chapman J.~M., Deacon R.~M., Redman M.~P., 2009, MNRAS, 397, 1386
\bibitem[\protect\citeauthoryear{Bl{\"o}cker}{1995}]{block95}
Bl{\"o}cker T., 1995, A\&A, 299, 755
\bibitem[\protect\citeauthoryear{Borkowski, Blondin \& Harrington}{1997}]{bor97} 
Borkowski K.~J., Blondin J.~M., Harrington J.~P., 1997, ApJL, 482, L97
\bibitem[\protect\citeauthoryear{Bollen et al.}{2021}]{boll21} 
Bollen D., Kamath D., Van Winckel H., De Marco O., Wardle M., 2021, MNRAS, 502, 445
\bibitem[\protect\citeauthoryear{Breen et al.}{2013}]{breen13}
Breen S.~L. Ellingsen S.~P., Contreras Y., Green J.~A., Caswell J.~L., Stevens J.~B., Dawson J.~R., Voronkov M.~A., 2013, MNRAS, 435, 524
\bibitem[\protect\citeauthoryear{Bujarrabal et al.}{1994}]{buj94}
Bujarrabal V., Alcolea J., Neri R., Grewing M., 1994, ApJL, 436, L169
\bibitem[\protect\citeauthoryear{Bujarrabal et al.}{1997}]{buj97} Bujarrabal V., Alcolea J., Neri R., Grewing M., 1997, A\&A, 320, 540
\bibitem[\protect\citeauthoryear{Casali et al.}{2007}]{cas07}
Casali M. et al., 2007, A\&A, 467, 777
\bibitem[\protect\citeauthoryear{Cohen et al.}{2004}]{coh04}
Cohen M., van Winckel H., Bond H.~E., Gull T.~R., 2004, AJ, I27, 2362
\bibitem[\protect\citeauthoryear{Caswell, Haynes, \& Goss}{Caswell et al.}{1980}]{cas80} 
Caswell J.~L., Haynes R.~F., Goss W.~M., 1980, AuJPh, 33, 639
\bibitem[\protect\citeauthoryear{Caswell et al.}{2010}]{cas10} 
Caswell J.~L. et al., 2010, MNRAS, 404, 1029
\bibitem[\protect\citeauthoryear{Caswell et al.}{2011}]{cas11} 
Caswell J.~L. et al., 2011, MNRAS, 417, 1964
\bibitem[\protect\citeauthoryear{Cerrigone et al.}{2017}]{cer17}
Cerrigone L., Umana G., Trigilio C., Leto P., Buemi C.~S., Ingallinera A., 2017, MNRAS, 468, 3450
\bibitem[\protect\citeauthoryear{Chapman}{1988}]{chap88}
Chapman J.~M., 1988, MNRAS, 230, 415
\bibitem[\protect\citeauthoryear{Chengalur et al.}{1993}]{chen93} 
Chengalur J.~N., Lewis B.~M., Eder J., Terzian Y., 1993, ApJS, 89, 189
\bibitem[\protect\citeauthoryear{Christianto \& Seaquist}{1998}]{chr98} 
Christianto H., Seaquist E.~R., 1998, AJ, 115, 2466
\bibitem[\protect\citeauthoryear{Condon et al.}{1998}]{con98} Condon J.~J., Cotton W.~D., Greisen E.~W., Yin Q.~F., Perley R.~A., Taylor G.~B., Broderick J.~J., 1998, AJ, 115, 1693
\bibitem[\protect\citeauthoryear{Davis, Seaquist \& Purton}{1979}]{dsp79}
Davis L.~E., Seaquist E.~R, Purton C.~R., 1979, ApJ, 230, 434
\bibitem[\protect\citeauthoryear{Dawson et al.}{2022}]{daw22}
Dawson J.~R. et al., 2022, MNRAS, 512, 3345
\bibitem[\protect\citeauthoryear{de Gregorio-Monsalvo et al.}{2004}]{deG04}
de Gregorio-Monsalvo I., G{\'o}mez Y., Anglada G., Cesaroni R., Miranda L.~F., G{\'o}mez J.~F, Torrelles J.~M., 2004, ApJ, 601, 921
\bibitem[\protect\citeauthoryear{Deacon, Chapman \& Green}{Deacon et al.}{2004}]{deac04}
Deacon R.~M., Chapman J.~M., Green A.~J., 2004, ApJS, 155, 595
\bibitem[\protect\citeauthoryear{Deacon et al.}{2007}]{deac07}
Deacon R.~M., Chapman J.~M., Green A.~J., Sevenster M.~N., 2007, ApJ, 658, 1096
\bibitem[\protect\citeauthoryear{Deguchi, Nakashima \& Balasubramanyam}{2001}]{deg01}
Deguchi S., Nakashima J., Balasubramanyam R., 2001, PASJ, 53, 305
\bibitem[\protect\citeauthoryear{Diamond et al.}{1994}]{dia94}
Diamond P.~J., Kemball A.~J., Junor W., Zensus A., Benson J., Dhawan V., 1994, ApJL, 430, L61
\bibitem[\protect\citeauthoryear{Elitzur}{1992}]{eli92}
Elitzur M., 1992, ARA\&A, 30, 75
\bibitem[\protect\citeauthoryear{Gaia Collaboration et al.}{2021}]{gaia21} Gaia Collaboration et al., 2021, A\&A, 649, A1
\bibitem[\protect\citeauthoryear{G{\'o}mez et al.}{2005}]{gom05}
G{\'o}mez J.~F., de Gregorio-Monsalvo I., Lovell J.~E.~J., Anglada G., Miranda L.~F., Su{\'a}rez O., Torrelles J.~M.,  G{\'o}mez Y., 2005, MNRAS, 364, 738
\bibitem[\protect\citeauthoryear{G{\'o}mez et al.}{2008}]{gom08}
G{\'o}mez J.~F., Su{\'a}rez O., G{\'o}mez Y., Miranda L.~F., Torrelles J.~M., Anglada G., Morata 0., 2008, AJ, 135, 2074
\bibitem[\protect\citeauthoryear{G{\'o}mez et al.}{2014}]{gom14} 
G{\'o}mez J.~F., Uscanga L., Su{\'a}rez O., Rizzo J.~R., de Gregorio-Monsalvo I., 2014, RMxAA, 50, 137
\bibitem[\protect\citeauthoryear{G{\'o}mez et al.}{2015}]{gom15}
G{\'o}mez J.~F. et al., 2015, ApJ, 799, 186
\bibitem[\protect\citeauthoryear{G{\'o}mez et al.}{2016}]{gom16}
G{\'o}mez J.~F., Uscanga L., Green J.~A., Miranda L.~F., Su{\'a}rez O., Bendjoya P., 2016, MNRAS, 461, 3259
\bibitem[\protect\citeauthoryear{G{\'o}mez et al.}{2017}]{gom17}
G{\'o}mez J.~F., Su{\'a}rez O., Rizzo  J.~R., Uscanga L., Miranda L.~F., Walsh A., Bendjoya P., 2017, MNRAS, 468, 2081
\bibitem[\protect\citeauthoryear{G{\'o}mez, Moran \& Rodr{\'i}guez}{1990}]{gom90}
G{\'o}mez Y., Moran J.~M., Rodr{\'i}guez L.~F., 1990, Rev. Mex. Astron. Astrofis., 20, 55
\bibitem[\protect\citeauthoryear{G{\'o}mez et al.}{2009}]{gom09}
G{\'o}mez Y., Tafoya D. Anglada G., Miranda L.~F., Torrelles J.~M., Patel N.~A., Franco Hern{\'a}ndez R., 2009, ApJ, 695, 930
\bibitem[\protect\citeauthoryear{Gordon et al.}{2021}]{vlass21} Gordon Y.~A. et al., 2021, ApJS, 257, 30
\bibitem[\protect\citeauthoryear{Green et al.}{2012}]{green12}
Green J.~A., McClure-Griffiths N.~M., Caswell J.~L., Robishaw T., Harvey-Smith L., 2012, MNRAS, 425, 2530
\bibitem[\protect\citeauthoryear{Groenewegen, van den Hoek \& de Jong}{1995}]{groene95}
Groenewegen M.~A.~T., van den Hoek L.~B., de Jong T., A\&A, 293, 381
\bibitem[\protect\citeauthoryear{Hambly et al.}{2008}]{ham08}
Hambly N.~C. et al., 2008, MNRAS, 384, 637
\bibitem[\protect\citeauthoryear{Hewett et al.}{2006}]{hew06}
Hewett P.~C., Warren S.~J., Leggett S.~K., Hodgkin S.~T., 2006, MNRAS, 367, 454
\bibitem[\protect\citeauthoryear{Ho \& Townes}{1983}]{ho83} 
Ho P.~T.~P., Townes C.~H., 1983, ARA\&A, 21, 239
\bibitem[\protect\citeauthoryear{Ivanova et al.}{2013}]{iva13}
Ivanova N. et al., 2013, A\&ARv, 21, 59
\bibitem[\protect\citeauthoryear{Ivezic, Nenkova \& Elitzur}{1999}]{ive99}
Ivezic Z., Nenkova M., Elitzur M., 1999, preprint (astro-ph/9910475)
\bibitem[\protect\citeauthoryear{Imai}{2007}]{imai07}
Imai H., 2007, in IAU Symp. 242, 279
\bibitem[\protect\citeauthoryear{Jura et al.}{1997}]{jur97} Jura M., Turner J., Balm, S.~P., 1997, ApJ, 474, 741. doi:10.1086/303476
\bibitem[\protect\citeauthoryear{Khouri et al.}{2021}]{kho21} Khouri T., Vlemmings W.~H.~T., Tafoya D., P{\'e}rez-S{\'a}nchez A.~F., S{\'a}nchez Contreras C., G{\'o}mez J.~F., Imai H., et al., 2021, NatAs, 6, 275
\bibitem[\protect\citeauthoryear{Knapp et al.}{1995}]{kna95} 
Knapp G.~R., Bowers P.~F., Young K., Phillips T.~G., 1995, ApJ, 455, 293
\bibitem[\protect\citeauthoryear{Kwok \& Feldman}{1981}]{kwo81a} 
Kwok S., Feldman P.~A., 1981, ApJL, 247, L67
\bibitem[\protect\citeauthoryear{Kwok, Purton, \& Keenan}{Kwok et al.}{1981}]{kwo81b} 
Kwok S., Purton C.~R., Keenan D.~W., 1981, ApJ, 250, 232
\bibitem[\protect\citeauthoryear{Lawrence et al.}{2007}]{law07}
Lawrence A. et al., 2007, MNRAS, 379, 1599
\bibitem[\protect\citeauthoryear{Lewis}{1989}]{lew89}
Lewis B.~M., 1989, ApJ, 338, 234
\bibitem[\protect\citeauthoryear{Lewis}{1997}]{lew97} 
Lewis B.~M., 1997, ApJS, 109, 489
\bibitem[\protect\citeauthoryear{te Lintel Hekkert et al.}{1991}]{telin91}
te Lintel Hekkert P., Caswell J.~L., Habing H.~J., Haynes R.~F., Norris R.~P., 1991, A\&AS, 90, 327
\bibitem[\protect\citeauthoryear{Manchado et al.}{1996}]{IAC96}
Manchado A., Guerrero M. A., Stanghellini L., Serra-Ricart M., 1996, The IAC morphological catalog of northern Galactic planetary nebulae. Instituto de Astrofisica de Canarias (IAC), La Laguna, Spain
\bibitem[\protect\citeauthoryear{Menten}{1991}]{men91} 
Menten K.~M., 1991, ApJL, 380, L75
\bibitem[\protect\citeauthoryear{Menten et al.}{2018}]{men18} Menten K.~M., Wyrowski F., Keller D., Kami{\'n}ski T., 2018, A\&A, 613, A49
\bibitem[\protect\citeauthoryear{Miller Bertolami}{2016}]{mil16}
Miller Bertolami  M.~M., 2016, A\&A, 588, A25
\bibitem[\protect\citeauthoryear{Miranda et al.}{2001}]{mir01}
Miranda L.~F., G{\'o}mez Y., Anglada G., Torrelles J.~M., 2001, Nature, 414, 
284
\bibitem[\protect\citeauthoryear{Miranda et al.}{2021}]{mir21} Miranda L.~F.,  et al., 2021, arXiv, arXiv:2105.05186
\bibitem[\protect\citeauthoryear{Monet et al.}{2003}]{mon03}
Monet D.~G. et al., 2003, AJ, 125, 984
\bibitem[\protect\citeauthoryear{Nyman, Hall \& Olofsson}{1997}]{nym97} 
Nyman L.~\AA, Hall P.~J., Olofsson H., 1998, A\&AS, 127, 185
\bibitem[\protect\citeauthoryear{Ohnaka et al.}{2013}]{ohn13} Ohnaka K., Boboltz D.~A., Mulitz-Schimel G., Izumiura H., Wittkowski M., 2013, A\&A, 559, A120
\bibitem[\protect\citeauthoryear{Olnon}{1975}]{ol75}
Olnon F.~M., 1975, A\&A, 39, 217
\bibitem[\protect\citeauthoryear{Panagia \& Felli}{1975}]{pan75}
Panagia N., Felli M., 1975, A\&A, 39, 1
\bibitem[\protect\citeauthoryear{P{\'e}rez-S{\'a}nchez et al.}{2017}]{per17} 
P{\'e}rez-S{\'a}nchez A.~F., Tafoya D., Garc{\'\i}a L{\'o}pez R., Vlemmings W.~H.~T., Rodr{\'\i}guez L.~F., 2017, A\&A, 601, A68
\bibitem[\protect\citeauthoryear{P{\'e}rez-S{\'a}nchez, Vlemmings \& Chapman}{2011}]{per11}
P{\'e}rez-S{\'a}nchez A.~F., Vlemmings W.~H.~T., Chapman J.~M., 2011, MNRAS, 418, L402
\bibitem[\protect\citeauthoryear{P{\'e}rez-S{\'a}nchez et al.}{2013}]{per13}
P{\'e}rez-S{\'a}nchez A.~F., Vlemmings W.~H.~T., Tafoya D., Chapman J.~M., 2013, MNRAS, 436, L79
\bibitem[\protect\citeauthoryear{Pottasch, Bignelly \& Zijlstra}{1987}]{pott87}
Pottasch S.~R., Bignelli C., Zijlstra A., 1987, A\&A, 177, L49
\bibitem[\protect\citeauthoryear{Purcell et al.}{2012}]{pur12} 
Purcell C.~R., Longmore S.~N., Walsh A.~J., Whiting M.~T., Breen S.~L., Britton T., Brooks K.~J., et al., 2012, MNRAS, 426, 1972 %doi:10.1111/j.1365-2966.2012.21800.x
\bibitem[\protect\citeauthoryear{Qiao et al.}{2016a}]{qia16a}
Qiao H.~H. et al., 2016a, ApJ, 817, 37
\bibitem[\protect\citeauthoryear{Qiao et al.}{2016b}]{qia16b}
Qiao H.~H. et al., 2016b, ApJS, 227, 26
\bibitem[\protect\citeauthoryear{Qiao et al.}{2018}]{qia18}
Qiao H.~H. et al., 2018, ApJS, 239, 15
\bibitem[\protect\citeauthoryear{Qiao et al.}{2020}]{qia20}
Qiao H.~H. et al., 2020, ApJS, 247, 5
\bibitem[\protect\citeauthoryear{Ramos-Larios et al.}{2009}]{ram09}
Ramos-Larios G., Guerrero M.~A., Su{\'a}rez O., Miranda L.~F., G{\'o}mez J.~F., 2009, A\&A, 501, 1207
\bibitem[\protect\citeauthoryear{Ramos-Larios et al.}{2012}]{ram12}
Ramos-Larios G., Guerrero M.~A., Su{\'a}rez O., Miranda L.~F., G{\'o}mez J.~F., 2012, A\&A, 545, A20
\bibitem[\protect\citeauthoryear{Reid}{1976}]{reid76}
Reid M.~J., 1976, ApJ, 207, 784
\bibitem[\protect\citeauthoryear{Reid \& Moran}{1981}]{reid81}
Reid, M.~J., Moran, J.~M., 1981, ARA\&A, 19, 231
\bibitem[\protect\citeauthoryear{Reid et al.}{1988}]{reid88}
Reid M.~J., Schneps M.~H., Moran, J.~M., Gwinn C.~R., Genzel R., Downes D., R{\"o}nn{\"a}ng B., 1988, ApJ, 330, 809
\bibitem[\protect\citeauthoryear{Reynolds}{1986}]{reyn86}
Reynolds S.~P., 1986, ApJ, 304, 713
\bibitem[\protect\citeauthoryear{Rizzo et al.}{2013}]{riz13} 
Rizzo J.~R., G{\'o}mez J.~F., Miranda L.~F., Osorio M., Su{\'a}rez O., Dur{\'a}n-Rojas M.~C., 2013, A\&A, 560, A82
\bibitem[\protect\citeauthoryear{Sahai \& Trauger}{1998}]{sah98}
Sahai R., Trauger J.~T., 1998, AJ, 116, 1357
\bibitem[\protect\citeauthoryear{Sahai, Morris \& Villar}{2011}]{sah11}
Sahai R., Morris M.~R., Villar G-~G., 2011, AJ, 141
\bibitem[\protect\citeauthoryear{S{\'a}nchez Contreras \& Sahai}{2012}]{sansa12}
S{\'a}nchez-Contreras C., Sahai R., 2012, ApJS, 203, 16
\bibitem[\protect\citeauthoryear{Sevenster}{2002}]{sev02}
Sevenster M.~N., 2002, AJ, 123, 2772
\bibitem[\protect\citeauthoryear{Sevenster et al.}{1997a}]{sev97a}
Sevenster M.~N., Chapman J.~M, Habing H.~J., Killeen N.~E.~B., Lindqvist M., 1997a, A\&AS, 122, 79
\bibitem[\protect\citeauthoryear{Sevenster et al.}{1997b}]{sev97b}
Sevenster M.~N., Chapman J.~M, Habing H.~J., Killeen N.~E.~B., Lindqvist M., 1997b, A\&AS, 124, 509
\bibitem[\protect\citeauthoryear{Sevenster et al.}{2001}]{sev01}
Sevenster M.~N., van Langevelde H.~J., Moody R.~A., Chapman J.~M, Habing H.~J., Killeen N.~E.~B., 2001, A\&A, 366, 481
\bibitem[\protect\citeauthoryear{Su{\'a}rez et al.}{2009}]{sua09} Su{\'a}rez O., G{\'o}mez J.~F., Miranda L.~F., Torrelles J.~M., G{\'o}mez Y., Anglada G., Morata O., 2009, A\&A, 505, 217
\bibitem[\protect\citeauthoryear{Su{\'a}rez et al.}{2015}]{sua15}
Su{\'a}rez O. et al., 2015, ApJ, 806, 105
\bibitem[\protect\citeauthoryear{Szymczak \& G{\'e}rard}{2004}]{szy04} 
Szymczak M., G{\'e}rard E., 2004, A\&A, 414, 235
\bibitem[\protect\citeauthoryear{Tafoya et al.}{2009}]{taf09}
Tafoya D., G{\'o}mez Y., Patel N.~A., Torrelles J.~M., G{\'o}mez J.~F., Anglada G., Miranda L.~F., de Gregorio-Monsalvo I., 2009, ApJ, 691, 611
\bibitem[\protect\citeauthoryear{Urquhart et al.}{2007}]{urqu07}
Urquhart J.~S., Busfield A.~L., Hoare M.~G., Lumsden S.~L., Clarke A.~J., Moore T.~J.~T., Mottram J.~C., Oudmaijer R.~D., 2007, A\&A, 461, 11
\bibitem[\protect\citeauthoryear{Uscanga et al.}{2008}]{usc08}
Uscanga L, G{\'o}mez Y, Raga A.~C., Cant{\'o} J., Anglada G., G{\'o}mez J.~F., Torrelles J.~M., Miranda L.~F., 2008, MNRAS, 390, 1127
\bibitem[\protect\citeauthoryear{Uscanga et al.}{2012}]{usc12}
Uscanga L, G{\'o}mez J.~F., Su{\'a}rez O., Miranda L.~F., 2012, A\&A, 547, A40
\bibitem[\protect\citeauthoryear{Uscanga et al.}{2014}]{usc14}
Uscanga L., G{\'o}mez J.~F., Miranda L.~F., Boumis P., Su{\'a}rez O., Torrelles J.~M., Anglada G., Tafoya D., 2014, MNRAS, 444, 217
\bibitem[\protect\citeauthoryear{van de Steene \& Jacoby}{2001}]{van01}
van de Steene G.~C., Jacoby G.~H., 2001, A\&A, 373, 536
\bibitem[\protect\citeauthoryear{Vassiliadis \& Wood}{1993}]{vass93}
Vassiliadis E., Wood P.~R., 1993, ApJ, 413, 641
\bibitem[\protect\citeauthoryear{Vassiliadis \& Wood}{1994}]{vass94}
Vassiliadis E., Wood P.~R., 1994, ApJS, 92, 125
\bibitem[\protect\citeauthoryear{Verdes-Montenegro et al.}{1989}]{ver89} 
Verdes-Montenegro L. et al., 1989, ApJ, 346, 193
\bibitem[\protect\citeauthoryear{Walsh et al.}{2009}]{wal09} Walsh A.~J., Breen S.~L., Bains I., Vlemmings W.~H.~T., 2009, MNRAS, 394, L70
\bibitem[\protect\citeauthoryear{Walsh et al.}{2011}]{wal11} 
Walsh A.~J. et al., 2011, MNRAS, 416, 1764
\bibitem[\protect\citeauthoryear{Walsh et al.}{2014}]{wal14} 
Walsh A.~J. et al., 2014, MNRAS, 442, 2240 
\bibitem[\protect\citeauthoryear{White, Becker \& Helfand}{White et al.}{2005}]{white05}
White R.~L., Becker R.~H., Helfand D.~J., 2005, AJ, 130, 586
\bibitem[\protect\citeauthoryear{Wood \& Churchwell}{1989}]{wc89}
Wood D.~O.~S, Churchwell E., 1989, ApJ, 340, 265
\bibitem[\protect\citeauthoryear{Zijlstra et al.}{2001}]{zij01}
Zijlstra A.~A., Chapman J.~M., te Lintel Hekkert P., Likkel L., Comeron F., Norris R.~P., Molster F.~J., R. J. Cohen, 2001, MNRAS, 322, 280
\bibitem[\protect\citeauthoryear{Zijlstra et al.}{1991}]{zij91} Zijlstra A.~A., Gaylard M.~J., te Lintel Hekkert P., Menzies J., Nyman L.-A., Schwarz H.~E., 1991, A\&A, 243, L9
\bibitem[\protect\citeauthoryear{Zijlstra et al.}{1989}]{zij89}
Zijlstra A.~A., te Lintel Hekkert P., Pottasch S.~R., Caswell J.~L., Ratag M., Habing, H.~J., 1989, A\&A, 217, 157
\end{thebibliography}
%\citep{iva13}
%\citep[e.g.,][]{san98,mor06,nor06,che17}


%%%%%%%%%%%%%%%%%%%%%%%%%%%%%%%%%%%%%%%%%%%%%%%%%%

%%%%%%%%%%%%%%%%% APPENDICES %%%%%%%%%%%%%%%%%%%%%



%%%%%%%%%%%%%%%%%%%%%%%%%%%%%%%%%%%%%%%%%%%%%%%%%%

% Don't change these lines
\bsp	% typesetting comment
\label{lastpage}
\end{document}